%% file: main.tex
\documentclass[aps,
pra,
superscriptaddress,
amsmath,
amssymb,
reprint,
floatfix]
{revtex4-2}
\usepackage[cm]{fullpage}
\usepackage{amsthm}
\usepackage[utf8]{inputenc}
\usepackage{times}
\usepackage{newtxmath}
\usepackage[colorlinks=true,linkcolor=blue, citecolor=magenta]{hyperref}
\usepackage{physics}
\usepackage{here}
\usepackage{mathtools}
\usepackage{bm}
\usepackage[english]{babel}
\usepackage{color}
\usepackage{comment}
\usepackage{ulem}
\usepackage[utf8]{inputenc}
\usepackage{color}

\usepackage{amsmath,amssymb,amsfonts,mathrsfs}

\usepackage{graphicx}
\newcommand{\mc}[1]{\mathcal{#1}}
\newcommand{\mr}[1]{\mathrm{#1}}

\newcommand{\mb}[1]{\mathbb{#1}}
\newcommand{\msf}[1]{\mathsf{#1}}

\newcommand{\what}[1]{\widehat{#1}}
\newcommand{\wtil}[1]{\widetilde{#1}}
\newcommand{\blue}[1]{\textcolor{blue}{#1}}
\newcommand{\rket}[1]{|#1)}

\theoremstyle{remark}

\begin{document}
	\title{
	Quantum many-body scars in bipartite Rydberg arrays originate from hidden projector embedding
	}
	\author{Keita Omiya}
	\affiliation{Department of Physics, ETH Z\"{u}rich, CH-8093 Z\"{u}rich, Switzerland}
	\affiliation{Laboratory for Theoretical and Computational Physics, Paul Scherrer Institute, Villigen CH-5232, Switzerland}
	\affiliation{Institute of Physics, Ecole Polytechnique F\'ed\'erale de Lausanne (EPFL), CH-1015 Lausanne, Switzerland}
	\author{Markus M\"uller}
	\affiliation{Laboratory for Theoretical and Computational Physics, Paul Scherrer Institute, Villigen CH-5232, Switzerland}
	\begin{abstract}
		We study the nature of the ergodicity-breaking "quantum many-body scar" states that appear in the PXP model describing constrained Rabi oscillations. For a {wide class of bipartite lattices} of Rydberg atoms, we reveal that the nearly energy-equidistant tower of these states arises from the Hamiltonian's close proximity to a generalized projector-embedding form~\cite{mori_shiraishi}, a  structure common to many models hosting quantum many-body scars. We construct a non-Hermitian, but strictly local extension of the PXP model hosting exact quantum scars, and show how various Hermitian scar-stabilizing extensions from the literature can be naturally understood within this framework. The  exact scar states are  obtained analytically as large spin states of explicitly constructed pseudospins. The quasi-periodic motion ensuing from the N\'eel state is finally shown to be the projection onto the Rydberg-constrained subspace of the precession of the large pseudospin.  
		
	\end{abstract}
\maketitle

\section{Introduction}
\input{introduction}

\section{Phenomenology of scar states}
\input{phenomenology}

\section{PXP model}
\input{setup}

\section{Analytical approximation for scar states}
\input{scar_state}

\section{Hamiltonian perturbations and generalized projector embedding formalism}
\input{perturbation}

\section{Summary and outlook}
\input{summary_outlook}

\textit{Acknowledgments.}---We acknowledge K. Pakrouski, S. Pappalardi and M. Sigrist for useful discussions. This work was supported by Grant No. 204801021001 of the SNSF.

\normalem
\bibliographystyle{unsrt}
\bibliography{reference}

\appendix
\input{appendix_new}

\end{document}

%% file: introduction.tex
The problem of ergodicity and thermalization has been a central issue in quantum statistical physics. In generic many-body systems, interactions are believed to entangle all degrees of freedom, pushing the dynamics toward a state of maximal entropy consistent with the global conservation laws. This intuition is at the root of the eigenstate thermalization hypothesis (ETH)~\cite{eth_deutsch,eth_srednicki,eth_review,Rigol2008}, which conjectures that for a small subsystem the reduced density matrix of a single many-body eigenstate is equivalent to that of the thermal density matrix. If true, this guarantees thermalization in systems with non-degenerate energy spectrum.  

Although the ETH is a plausible and rather generic scenario ensuring thermalization, various non-ergodic systems which violate the ETH have been reported. Probably the only  class to be robust against generic small perturbations consists in many-body localized (MBL) systems featuring strong quenched disorder that suppresses the mixing mediated by local interactions~\cite{mbl_gornyi,mbl_basko,mbl_review}. At least in locally finite, one-dimensional lattice systems MBL is believed to occur as a genuine phase of matter~\cite{Imbrie2016}. 

Another class of systems that violate the ETH in more subtle terms was recently discovered, as initiated by the experimental observation ~\cite{qmbs_experiment}
 of anomalously long-lived quantum revivals in the dynamics of a chain of Rydberg atoms starting from a density wave state. In particular the lifetime of these oscillations was significantly longer than the typical dynamical time scale. Many of the eigenstates that play a dominant role in that particular  dynamical trajectory turn out to be far from thermal, and moreover, come with nearly equidistant eigenenergies~\cite{Turner2018,turner_tower}. In analogy to similar phenomena observed in single particle billiards, this phenomenology was referred to as a quantum many-body scar (QMBS)~\cite{Turner2018, qmbs_review}. 

These findings motivated the study of scar states in various non-integrable models where exact towers of exceptional states with equidistant spectrum were identified. By now many models of spins~\cite{tos_aklt,tos_xy,tos_exact_2,Lin2020} as well as fermions~\cite{Kiryl2020,Kiryl2021} have been found to host scar states. They can be constructed analytically and thus help us understand QMBS from a unified perspective. Indeed, many of these models share a common algebraic structure, which allows for a systematic construction of models hosting QMBS~\cite{mori_shiraishi, Kiryl2020, tos_aklt_unified, ODea2020TunnelsToTowers, Ren2021Quasisymmetry}.
For example, Ref.~\cite{tos_aklt_unified} put forward that scar states can often be generated from a simple reference state (such as a ferromagnetic state) by acting on it repeatedly with a ladder operator, which is part of a "spectrum-generating algebra" (SGA)~\cite{eta_pair_rsga}. 


\begin{figure}
    \centering
    \includegraphics[width=.48\textwidth]{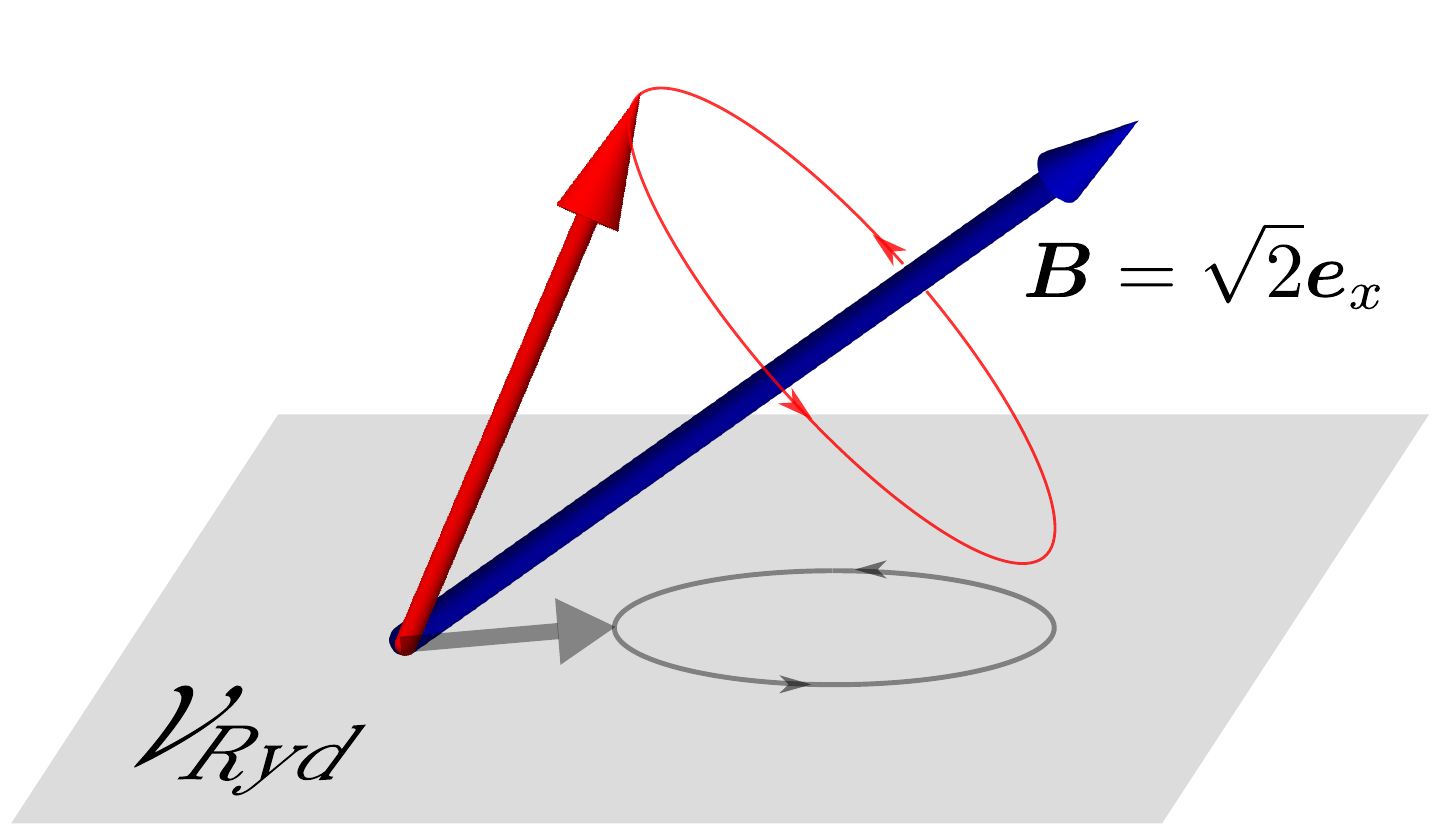}
    \caption{Pictorial illustration of the dynamics of the PXP model and its scar-enhanced versions on bipartite lattices: Blocks of two neighboring Rydberg constrained spins $S=1/2$ from different sublattices form effective $S=1$ degrees of freedom and combine to a maximal total spin state represented by the red arrow. That large spin precesses around the $x$-axis. The actual dynamics of the system follow the projection of the large spin onto the constrained subspace $\mc{V}_{\mr{Ryd}}$. The ETH-violating scar states correspond to projections of specific states of the large spin. The projected motion as well as the scar states turn out to be {\em independent} of the dimerization chosen to construct the large spin.
    }
    \label{fig:illustration}
\end{figure}

While the above structures 
may explain the occurrence of scar states in many models, their potential connection with the first and experimentally most relevant QMBS system - the Rydberg chain and the associated PXP model - has remained unclear. A main obstacle has been the fact that the QMBS phenomenology in the Rydberg chains is only approximate, with nearly but not exactly equidistant eigenenergies in the tower of exceptional states.
In the absence of a thorough understanding of the origin of those scars in the PXP model, numerous theoretical approaches have been suggested, including the emergence of an $SU(2)$ algebra~\cite{nearly_su2}, a $\pi$-magnon condensation~\cite{Iadecola2019pimagnon}, an approximate description of scar states by matrix product states 
~\cite{Lin2019mps}, or the proximity to an integrable point~\cite{Khemani2019integral}. While these approaches succeeded in capturing the scar states numerically, many important questions remained. Neither the subspace spanned by scar states nor the structure of individual scar states has been identified, and in particular there is no compact analytical form 
for the latter. Although Refs.~\cite{nearly_su2, Iadecola2019pimagnon} have suggested possible algebraic structures underlying the approximate tower of states, those do not fully elucidate how the tower of states arises and how these structures are related to other known models hosting exact scars. Furthermore, while it is known that certain modifications of the PXP model can strongly stabilize dynamic revivals from the initial density wave (N\'eel) state~\cite{Kieran2020}, it is not well understood why the proposed perturbation schemes actually converge and what is the structure of the scar states they stabilize. To date there are no fully satisfactory answers to the above open questions, not even for the intensely studied one-dimensional PXP model. And yet much less is understood for higher-dimensional analogues, some of which were studied experimentally in the form of PXP models on various two dimensional, bipartite lattices, cf. Ref.~\cite{Bluvstein1355}. Empirically it was found that  the dynamics starting from the N\'eel state exhibits quasi-periodic motion over time scales that vary significantly among the different lattices, but a theory rationalizing this
observation has been missing.  

In this {paper} we shed new light on these open questions. In particular, we provide explicit analytical wavefunctions that are excellent approximations for the scar states of the PXP model {on bipartite lattices}, and come even closer to the scar states of some of the scar-enhancing modifications of PXP Hamiltonians. In simple cases, such as a Rydberg chain, 
these wavefunctions can be seen as Rydberg projections  of a precessing large spin state that we construct {\em explicitly} from the elementary degrees of freedom. 
However, on more complex lattices where analogous wavefunctions still capture the empirical scar states well, they transcend the simple picture of a hidden large spin in an effective field, suggesting that the ultimate description of scar states should be more general. Instead, our construction suggests that bipartite lattices with larger coordination numbers tend to be farther from hosting a tower of exactly equidistant scar states.
We further discuss a generic structure of Hamiltonians shared by almost all QMBS-hosting models known so far, and reveal how the PXP model fits into a generalization of this framework. 

%% file: phenomenology.tex
\label{sec:phenomenology}
The existence of scar states, nonthermal eigenstates in systems with a non-integrable Hamiltonian, is a consequence of a structure common to almost all models with analytically known towers of exact scar states~\cite{tos_aklt_unified, KO2022}
: The Hamiltonian $H$ can be split into two parts, $H=H_{\mr{spec}}+H_{\mr{ann}}$. The scar states, denoted by $\ket{\mc{S}_n}$, are eigenstates of $H_{\mr{spec}}$ with equidistant eigenenergies 
$E_n$ [i.e., $(E_n-E_m)/\Omega \in \mathbf Z$],  
and are annihilated by $H_{\mr{ann}}$ ($H_{\mr{ann}}\ket{\mc{S}_n}=0$). Both $H_{\mr{spec}}$ and $H_{\mr{ann}}$ are local in the sense that they are sums of local operators. Any superposition of $\ket{\mc{S}_n}$ exhibits exact revivals under unitary time evolution by multiples of $1/\Omega$ ($\hbar =1$). $H_{\mr{spec}}$ {is integrable and }often consists in a simple Zeeman term
~\cite{tos_xy, tos_exact_2}. 
$H_{\mr{ann}}$ renders the remainder of the spectrum chaotic. In most cases it can be further decomposed into scar-annihilating {\em local} operators. Sets of such local annihilators that have a non-trivial common kernel (namely the prospective scar states) can be systematically constructed~\cite{mori_shiraishi, Kiryl2020, Kiryl2021, Ren2021Quasisymmetry}. An instructive example is the projector-embedding by Shiraishi and Mori~\cite{mori_shiraishi}.   
{In this approach, $H_{\mr{ann}}$ is written as $H_{\mr{ann}}=\sum_i\mc{P}_ih_i\mc{P}_i$, where $\mc{P}_i$ is a local projector and $h_i$ is any local operator. Scar states are annihilated separately by every $\mc{P}_i$. For various models, such a decomposition of the Hamiltonian has been identified, although sometimes one has to enlarge the Hilbert space to find local annihilators~\cite{KO2022}.} 
{However, previous attempts~\cite{nearly_su2,Iadecola2019pimagnon,Lin2019mps,Khemani2019integral} to understand the scar states in the PXP model have not succeeded to uncover a similar structure in the Hamiltonian, which has hindered a unified understanding of the nature of the QMBS appearing in PXP models and of the way in which they relate to the analytically known QMBS in other systems.} Here, we will close this gap by showing how the {PXP model} fits within such a framework {: It can be approximately expressed in a \textit{generalized} projector-embedding form,}
\begin{equation}
    {HP_{\mr{Ryd}}{\ket*{\psi}}\approx P_{\mr{Ryd}}\left(H_{\mr{Z}}+\sum_ih_i\mc{P}_i\right){\ket*{\psi}},}
\end{equation}
{where $P_{\mr{Ryd}}$ imposes the Rydberg blockade condition that prohibits simultaneous excitations of neighboring atoms, a precise definition being given further below in Eq.~\eqref{eq:projector}. $H_{\mr{Z}}$ turns out to be a simple Zeeman term, while $H_{\mr{ann}}=\sum_ih_i\mc{P}_i$ annihilates states of  maximal total spin, once $H$ is expressed in a basis of $S=1$ ``block-spins", each of which describes the Rydberg-allowed configurations of two neighboring $S=1/2$ spins.} Note that the operator on the right-hand side (RHS) acts on the partially restricted 
Hilbert space of {block spins} and is non-Hermitian, while the Hamiltonian on the left-hand side (LHS) acts essentially in the space that satisfies the Rydberg constraint. Such a non-Hermiticity often arises when the action of a Hamiltonian is lifted to a Hilbert space of larger dimension. Indeed a similar structure, arises in the AKLT model when one tries to understand its scar states in terms of local annihilators ~\cite{KO2022}.

%% file: setup.tex
\label{sec:setup}
\subsection{Hamiltonian}
\begin{figure}
    \centering
    \includegraphics[width=.3\textwidth]{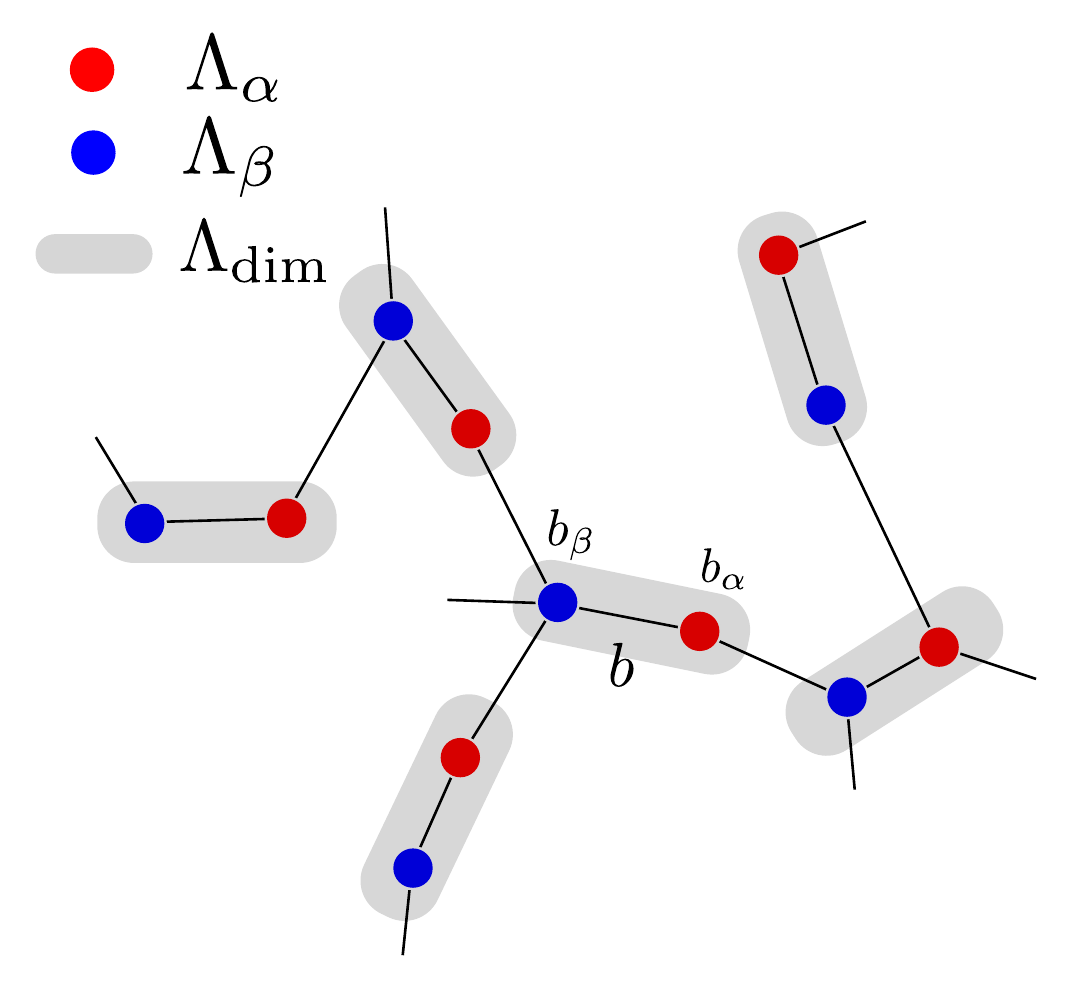}
    \caption{{Graphical representation of a bipartite lattice structure with its two sublattices $\Lambda_\alpha$ and $\Lambda_\beta$, colored in blue and red. The grey bars define a choice for a lattice of dimers, $\Lambda_{\mr{B}}$, each hosting a block spin.}}
    \label{fig:lattice}
\end{figure}
To describe a chain of Rydberg atoms with strong nearest neighbor repulsion between excited atoms Turner et al.~\cite{Turner2018} introduced the PXP model on a one-dimensional lattice. {Here we study this model and its generalization to a large class of bipartite lattices in arbitrary dimensions.}

{We consider an undirected graph $G=(\Lambda, E)$, where $\Lambda$ is a set of lattice sites (or vertices) and $E\subset\Lambda\times\Lambda$ is a set of edges labelled by their non-ordered endpoints, $(i,j) = (j,i) \in E$. Since $\Lambda$ is bipartite, there are two disjoint sublattices $\Lambda_\alpha$ and $\Lambda_\beta$ that cover the whole lattice ($\Lambda_\alpha\cup\Lambda_\beta=\Lambda$), and $(i,j)\in E$ implies either $i\in\Lambda_\alpha, j\in\Lambda_\beta$ or $i\in\Lambda_\beta, j\in\Lambda_\alpha$. Throughout most of this paper we assume $|\Lambda_\alpha|=|\Lambda_\beta|\equiv N_b$ (and thus $|\Lambda|=2N_b$). The PXP model on $\Lambda$ is defined as}
\begin{equation}\label{eq:pxp PBC}
	H=\sum_{i\in\Lambda}\left({X_i\prod_{j\,| (i,j)\in E}P_j}\right),
\end{equation} 
where $P\coloneqq\dyad{\downarrow}$ projects onto non-excited atoms while $X\coloneqq\dyad{\uparrow}{\downarrow}+\dyad{\downarrow}{\uparrow}$ drives the transition between ground and excited states. 
This Hamiltonian describes Rabi oscillations of individual atoms subject to the ``Rydberg" constraint  which requires  their neighboring atoms to be in their ground state, as implemented by the projectors $P$.

{We assume that $H$ is left invariant by the action of a crystalline symmetry group $G$, represented by operators $O_g$ that commute with $H$,  $[H, O_g]=0\,\forall g\in G$.}
{In the following, we focus on two examples: the one dimensional chain and the honeycomb lattice.}

\subsubsection{{One-dimensional chain}}
{For a one-dimensional chain, $\Lambda=\{1,\cdots,2{N_b}\}$ and $E=\{(i,i+1);\,i\in\Lambda\}$ with periodic boundary conditions ($i+2N_b\equiv i$). The Hamiltonian then becomes 
} 
\begin{equation}
    {H=\sum_{i\in\Lambda}P_{i-1}X_iP_{i+1}.}
\end{equation}
{In this case, the sublattices consist in the odd, $\Lambda_\alpha=\{2k+1;0\leq k\leq N_b-1\}$, and the even sites, $\Lambda_\beta=\{2k;1\leq k\leq N_b\}$, respectively. This model is  invariant under translation by one lattice spacing, which we denote by $T$, i.e., $[H,T]=0$. {The full lattice symmetry group also contains rotations around axes perpendicular to the chain, but those will not be of importance for our discussion.}} 

\subsubsection{{Honeycomb lattice}}
\begin{figure}
    \centering
    \includegraphics[width=.3\textwidth]{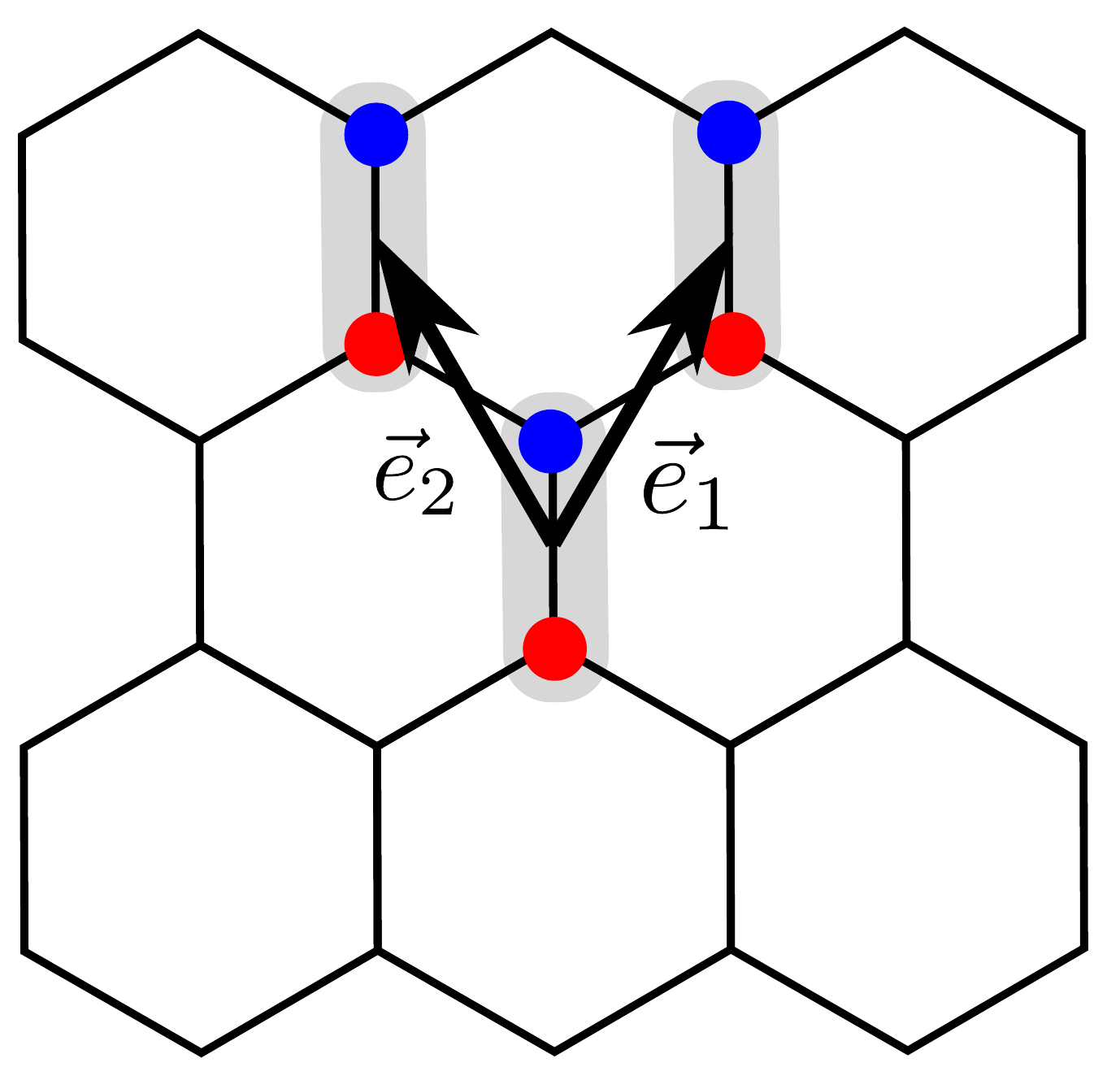}
    \caption{Honeycomb lattice with its two sublattices $\Lambda_\alpha$ and $\Lambda_\beta$ coloerd in red and blue, respectively. The black arrows indicate $\Vec{e}_{1,2}$, the two basis vectors of the lattice used in the text.}
    \label{fig:honeycomb}
\end{figure}
{Another bipartite lattice of interest is the honeycomb lattice. To define the lattice $\Lambda$, it is easiest to directly define the sublattices $\Lambda_\alpha$ and $\Lambda_\beta$, namely, $\Lambda_\alpha=\{\Vec{r}=a_1\Vec{e}_1+a_2\Vec{e}_2;a_1,a_2\in\mb{Z}\}$ and $\Lambda_\beta=\{\Vec{r}+\Vec{e}_y;\Vec{r}\in\Lambda_\alpha\}$ where $\Vec{e}_{1,2}\coloneqq \pm\sqrt{3}/2\Vec{e}_x +\,3/2\Vec{e}_y$ are a set of basis vectors of $\Lambda_\alpha$ (see Fig.~\ref{fig:honeycomb}). We also implement PBC's, identifying $\Vec{r}$ and $\Vec{r}+N_i\Vec{e}_i$ for $i=1,2$ with some positive integers $N_i$. The PXP model on the honeycomb lattice is then defined as}
\begin{equation}\begin{split}
    {H}&{=\sum_{\Vec{r}\in\Lambda}X_{\Vec{r}}\prod_{\vec{r}'\in\partial\vec{r}}P_{\Vec{r}'},}
    \end{split}
\end{equation}
{where $\partial\vec{r}\coloneqq\{\vec{r}'\in\Lambda|(\vec{r},\vec{r}')\in E\}$.}
{Note that the honeycomb lattice  is invariant under a site-centered rotation by an angle $2\pi/3$.}

\subsection{Dimerization}
The Hamiltonian {Eq.~\eqref{eq:pxp PBC}} commutes with the global projection operator $P_{\mr{Ryd}}\coloneqq\prod_{{(i,j)\in E}}(1-\dyad{\uparrow\uparrow})_{i,{j}}$ which prohibits neighboring atoms from being simultaneously excited. In the sequel we focus on the corresponding constrained subspace $\mc{V}_{\mr{Ryd}}\equiv P_{\mr{Ryd}}\mb{C}^{2\otimes2N_b}$. 
The constraint bars two neighboring spins from being up simultaneously. Hence, for the spins on an edge $b=(b_\alpha,b_\beta)$ {with neighboring sites $b_\alpha\in\Lambda_\alpha$ and $b_\beta\in\Lambda_\beta$},
only three configurations are possible.
Similarly as in Ref.~\cite{Lin2019mps} we identify them
with the states of an artificial $S=1$ {``block-spin"} by defining {$\ket{0}_b\coloneqq\ket{\downarrow\downarrow}_{b_\alpha,b_\beta}, \ket{+}_b\coloneqq\ket{\downarrow\uparrow}_{b_\alpha,b_\beta}$, and $\ket{-}_b\coloneqq\ket{\uparrow\downarrow}_{b_\alpha,b_\beta}$},
where $\ket{\pm}$ and $\ket{0}$ are eigenstates of the corresponding $S^z$-operator with eigenvalues $\pm1$ and $0$, respectively. 


{{Let us now consider a dimerization of the lattice, i.e., a covering of all lattice sites with a  set $\Lambda_{\mr{B}}$ of disjoint bonds $b$ (see Fig.~\ref{fig:lattice}). In general this choice is not unique. }
 We then write $b\equiv(i,j)\in\Lambda_{\mr{B}}$, and $b_\alpha\equiv i\,(b_\beta\equiv j)$ if $i\in\Lambda_\alpha\,(j\in\Lambda_\beta)$. For any particular choice of the dimer-covering $\Lambda_{\mr{B}}$, the PXP Hamiltonian can be expressed in the} associated $S=1$ basis as 
\begin{equation}\label{eq:decomposition H}
    \begin{split}
    {H=H_{\mr{Z}}+H_1+H_2,}
    \end{split}
\end{equation}
{where}
\begin{equation}\label{eq:pxp block-basis}
    \begin{split}
    {H_{\mr{Z}}}&{\coloneqq\sqrt{2}\sum_{b\in\Lambda_{\mr{B}}}S_b^x,}\\
    {H_1}&{\coloneqq\sum_{b\in\Lambda_{\mr{B}}}\dyad{0}{+}_b\left(\prod_{\substack{c\in\Lambda_{\mr{B}}\\c_\beta\in\partial b_\alpha\setminus\{b_\beta\}}}\left(1-\dyad*{-}\right)_c-1\right)}\\
    &{+\sum_{b\in\Lambda_{\mr{B}}}\dyad*{0}{-}_b\left(\prod_{\substack{d\in\Lambda_{\mr{B}}\\d_\alpha\in\partial b_\beta\setminus\{b_\alpha\}}}\left(1-\dyad*{+}\right)_d-1\right)}
    \end{split}
\end{equation}
and $H_2=H_1^\dag$. 
We now view $H_1$ and $H_2$ as perturbations of the leading Zeeman term $H_Z$, which acts on the set of block spins of $S=1$. {Later we will find it useful to write the product over projectors as a sum over increasingly many neighbors and express $H_1$ in the following way,}
\begin{equation}\label{eq:H1 expanded}\begin{split}
    {H_1}&{=\sum_{b\in\Lambda_{\mr{B}}}\left(\sum_{k=1}^{z_{b_\alpha}-1}(-1)^kh^{k;\alpha}_b+\sum_{k=1}^{z_{b_\beta}-1}(-1)^kh^{k;\beta}_b\right)}\\
    {h^{k;\alpha}_b}&{=\sum_{\substack{\{c_j\}_{j=1}^k\subset\Lambda_{\mr{B}}\\(c_j)_\beta\in\partial b_\alpha\setminus\{b_\beta\}}}\dyad*{0,-,\cdots,-}{+,-,\cdots,-}_{b,c_1,\cdots,c_k}}\\
    {h^{k;\beta}_b}&{=\sum_{\substack{\{d_j\}_{j=1}^k\subset\Lambda_{\mr{B}}\\(d_j)_\alpha\in\partial b_\beta\setminus\{b_\alpha\}}}\dyad*{0,+,\cdots,+}{-,+,\cdots,+}_{b,d_1,\cdots,d_k},}
    \end{split}
\end{equation}
{where $z_i$ is the coordination number of the site $i$.}
{
At the level of block spins the Rydberg constraint only has to be imposed between adjacent block spins and takes the following form,} 
\begin{equation}\label{eq:projector}
    {P_{\mr{Ryd}}=\prod_{(b_\beta,b'_\alpha)\in E}\left(1-\dyad*{+,-}\right)_{b,b'}.}
\end{equation}
{$P_{\mr{Ryd}}$ can now be seen as a projecting map from the spin-$1$ space $\mb{C}^{3\otimes N_b}$ to the Rydberg-constrained subspace $\mc{V}_{\mr{Ryd}}$.}

{Note that in general, a dimer covering breaks some lattice symmetries of the original model, and thus some of the symmetries of the original $S=1/2$ model cannot, {or at least not easily}, be expressed at the level of the $S=1$ representation, such that the $S=1$ Hamiltonian only possesses a reduced set of symmetries.} {We will address this issue in Sec.~\ref{sec:analytical approximation}, where we will see that the trial wavefunctions of the scar states that we will construct from specific dimerizations do not depend on that choice.} 

\subsubsection{{PXP chain}}
{For the one-dimensional PXP model with PBC, we may choose the dimer-covering $\Lambda_{\mr{B}}\coloneqq\{(2b-1,2b);\,1\leq b\leq N_b\}$. With this choice, the "perturbations" read}
\begin{equation}\begin{split}
    {H_1}&{=-\sum_{b\in\Lambda_{\mr{B}}}\left(\ket*{+,0}+\ket*{0,-}\right)\bra*{+,-}_{b,b+1}}.
    \end{split}
\end{equation}
{Note that when expressed in the block spin ($S=1$) basis, the model is only explicitly invariant with respect to translations by a full block spin or dimer, which corresponds to two $S=1/2$ sites, i.e., $T^2$. The invariance under $T$ is instead hidden in this basis.}

It is worth noting that with $H$ written in the form of Eqs.~\eqref{eq:decomposition H}{-\eqref{eq:H1 expanded}}  one can elegantly recover the  exact non-thermal zero energy eigenstate of {the 1D PXP model} that was found in Ref.~\cite{Lin2019mps}, as we show in {Appendix.}~\ref{app:ML state}. 

Numerical studies have established that the PXP chain hosts a set of ${2N_b}+1$ nonthermal eigenstates, $\ket*{\mc{S}^{\mr{PXP}}_n}$, that have high overlap with the N\'eel state $\ket{\mb{Z}_2}\coloneqq\bigotimes_{b\in\Lambda_{\mr{B}}}\ket{\downarrow\uparrow}_{{2b-1,2b}}=\bigotimes_{b\in\Lambda_{\mr{B}}}\ket{+}_b$. Moreover, these states have {almost equidistant} energies $E_n-E_{n-1}\approx \Omega_{\mr{PXP}}=1.33$ in the middle of the spectrum, while their spacing decreases towards the tails of the spectrum, (e.g. ${E_{N_b}-E_{N_b-1}}\approx0.968$ for ${N_b=10}$ block spins). 
However, the question as to what characterizes these scar states and how the subspace they span relates to a decomposition of the Hamiltonian as in  Eq.~\eqref{eq:pxp block-basis} has remained open.

\subsubsection{{Honeycomb lattice}}
{A natural dimer covering for the honeycomb lattice is $\Lambda_{\mr{B}}\coloneqq\{\Vec{R}\equiv(\Vec{r},\Vec{r}+\Vec{e}_y);\Vec{r}\in\Lambda_\alpha\}$. In this case, the perturbations become}
\begin{equation}\begin{split}
    H_1=&-\sum_{\Vec{R}\in\Lambda_{\mr{B}}}\sum_{i=1,2}\left(\ket*{+,0}+\ket*{0,-}\right)\bra*{+,-}_{\Vec{R},\Vec{R}+\Vec{e}_i}\\
    &+\sum_{\Vec{R}\in\Lambda_{\mr{B}}}\dyad*{0,-,-}{+,-,-}_{\Vec{R},\Vec{R}+\Vec{e}_1,\Vec{R}+\Vec{e}_2}\\
    &+\sum_{\Vec{R}\in\Lambda_{\mr{B}}}\dyad*{+,+,0}{+,+,-}_{\Vec{R}-\Vec{e}_1,\Vec{R}-\Vec{e}_2,\Vec{R}}.
    \end{split}
\end{equation}
By exact diagonalization of this model with PBC's (with $N_1 = N_2=3 $ unit cells in the directions $\vec{e}_{1,2}$, respectively), we found that it also hosts a similar set of $2N_b+1$ nonthermal eigenstates $\ket*{\mc{S}^{\mr{PXP}}_n}$ that have high overlap with the N\'eel state $\ket*{\mb{Z}_2}\coloneqq\bigotimes_{\Vec{r}\in\Lambda_\alpha}\ket*{\downarrow\uparrow}_{\Vec{r},\Vec{r}+\Vec{e}_y}=\bigotimes_{\Vec{R}\in\Lambda_{\mr{B}}}\ket*{+}_{\Vec{R}}$. 
Like in the 1D case, the spin-1 formulation of this model of a specific dimerization does not reflect the site-centered rotational symmetry anymore.

We will see {later} that neither the Zeeman term $H_{\mr{Z}}$, nor $H_1+H_2$ correspond directly to the parts $H_{\mr{spec}}$ and $H_{\mr{ann}}$ of a standard decomposition of $H$ acting on the Rydberg-constrained Hilbert space. However, the Zeeman term and the closeness of its magnitude $\sqrt{2}$ to $\Omega_{\mr{PXP}}$ 
suggest that the relevant Hilbert space to consider is the full $S=1$ sector $\mb{C}^{3\otimes N}$ of the chain, rather than merely its smaller subspace $\mc{V}_{\mr{Ryd}}$.
Moreover, in order to establish a connection of the Rydberg system to other models with exactly known scar states, the structure found there needs to be generalized to a system with constraints.

%% file: scar_state.tex
\label{sec:analytical approximation}
{We now proceed with the Hamiltonian written in a $S=1$ basis associated with a certain dimer covering. While the constructions we will make thus explicitly depend on the chosen covering, we will eventually find, rather remarkably, that the resulting trial wavefunctions are in fact independent of that choice. }

We now assume that $H_{\mr{Z}}$ in Eq.~\eqref{eq:pxp block-basis} essentially takes the role of $H_{\rm spec}$. It is then natural to consider its eigenstates $\ket*{\wtil{S}_n}\coloneqq(J^-)^{N-n}\bigotimes_{b\in\Lambda_{\mr{B}}}\ket*{\what{+}}_b$ as approximate trial wavefunctions for $\ket*{\mc{S}_n^{\mr{PXP}}}$, where $\ket*{\what{\pm}}$ and $\ket*{\what{0}}$ constitute the spin-$1$ basis diagonalizing $S^x$ (rather than $S^z$), with eigenvalues $\pm1$ and $0$, respectively, and $J^\pm\coloneqq\mp i\sum_{b\in\Lambda_{\mr{B}}}\what{S}_b^\pm$ with $\what{S}_b^\pm\coloneqq S_b^y\pm iS^z_b$ are collective ladder operators.
However, as the $\ket*{\wtil{S}_n}$ do not satisfy the Rydberg constraint, we instead consider the following projections $\ket{S_n}$:  
\begin{equation}\label{eq:trial state}
    \ket*{S_n}\coloneqq P_{\mr{Ryd}}\left[\left(J^-\right)^{N-n}\bigotimes_{b\in\Lambda_{\mr{B}}}\ket*{\what{+}}_b\right]
    =P_{\mr{Ryd}}\ket*{\wtil{S}_n}.
\end{equation}
Note that $\ket*{\wtil{S}_n}$ is an eigenstate of $H_{\mr{Z}}$ with eigenvalue $n\sqrt{2}$, and has maximal total pseudospin. 
Interestingly, the N\'eel state can be written as an exact superposition of these states $\ket*{S_n}$, 
\begin{equation}
\label{eq:Neel}
\ket*{\mb{Z}_2}=\bigotimes_{b\in\Lambda_{\mr{B}}}\ket*{{+}}_b=
\frac{1}{2^N}\sum_{k=0}^{2N}
\frac{1}{k!}P_{\mr{Ryd}}\left(J^-\right)^k\bigotimes_{b\in\Lambda_{\mr{B}}}\ket*{\what{+}}_b,
\end{equation}
where we used $P_{\mr{Ryd}}\ket*{\mb{Z}_2}=\ket*{\mb{Z}_2}$ and $\ket*{+}=\frac{1}{2}e^{{i}
\what{S}^-}\ket*{\what{+}}$\blue{.} An equivalent relation holds upon replacing $\ket*{\what{+}}$ and $\what{S}^-$ with $\ket*{\what{-}}$ and $-\what{S}^+$, respectively. 

{As we pointed out above, the choice of a particular dimer covering and the restriction of the Hamiltonian to its Zeeman part $H_Z$  } break {some of the symmetries of the original $S=1/2$ system}. This is then imprinted on the parent states $\ket*{\wtil{S}_n}$, too. Interestingly, however, the Rydberg projection restores {the broken symmetries} in $\ket*{{S}_n}$. This in turn implies that one obtains the same states $\ket*{{S}_n}$ from {\em any} choice of dimerization pattern. {Indeed, we can show that $\ket*{S_n}$ can be written in a manifestly invariant manner,}
\begin{equation}\label{eq:|Sn> inv}
    {\ket*{S_n}\propto P_{\mr{Ryd}}U_\varphi\left(-\sum_{i\in\Lambda_\alpha}\sigma_i^-+\sum_{i'\in\Lambda_\beta}\sigma_{i'}^-\right)^{N_b-n}\bigotimes_{j\in\Lambda}\ket*{\uparrow}_j,}
\end{equation}
{where $\sigma^-\coloneqq\dyad*{\downarrow}{\uparrow}$ is the lowering operator of the original $S=1/2$ operators, and $U_\varphi = \prod_i U_i^\varphi$ is a product over all sites of the same single-site operator, $U_i^\varphi$. A detailed derivation of the representation (\ref{eq:|Sn> inv}) is given in Appendix.~\ref{sec:invariance}. In the 1D case, the operator in the parantheses of Eq.~\eqref{eq:|Sn> inv} can be interpreted to mean that individual excitations in the tower of states carry momentum $\pi$. {On the honeycomb lattice instead the excitations can be interpreted as carrying momentum $K$ corresponding to the corners of the first Brillouin zone.}} 

{For an arbitrary bipartite lattice, every} dimerization provides a different, though ultimately equivalent description for the PXP dynamics. It lifts it from the Rydberg-constrained subspace to the full Hilbert space, where the dynamics reduces mainly to the precession of a large spin. The embedding of the large spin in the full Hilbert space depends explicitly on the chosen dimer covering, but its projection onto the Rydberg subspace does not.

\begin{figure}
    \centering
    \includegraphics[width=0.48\textwidth]{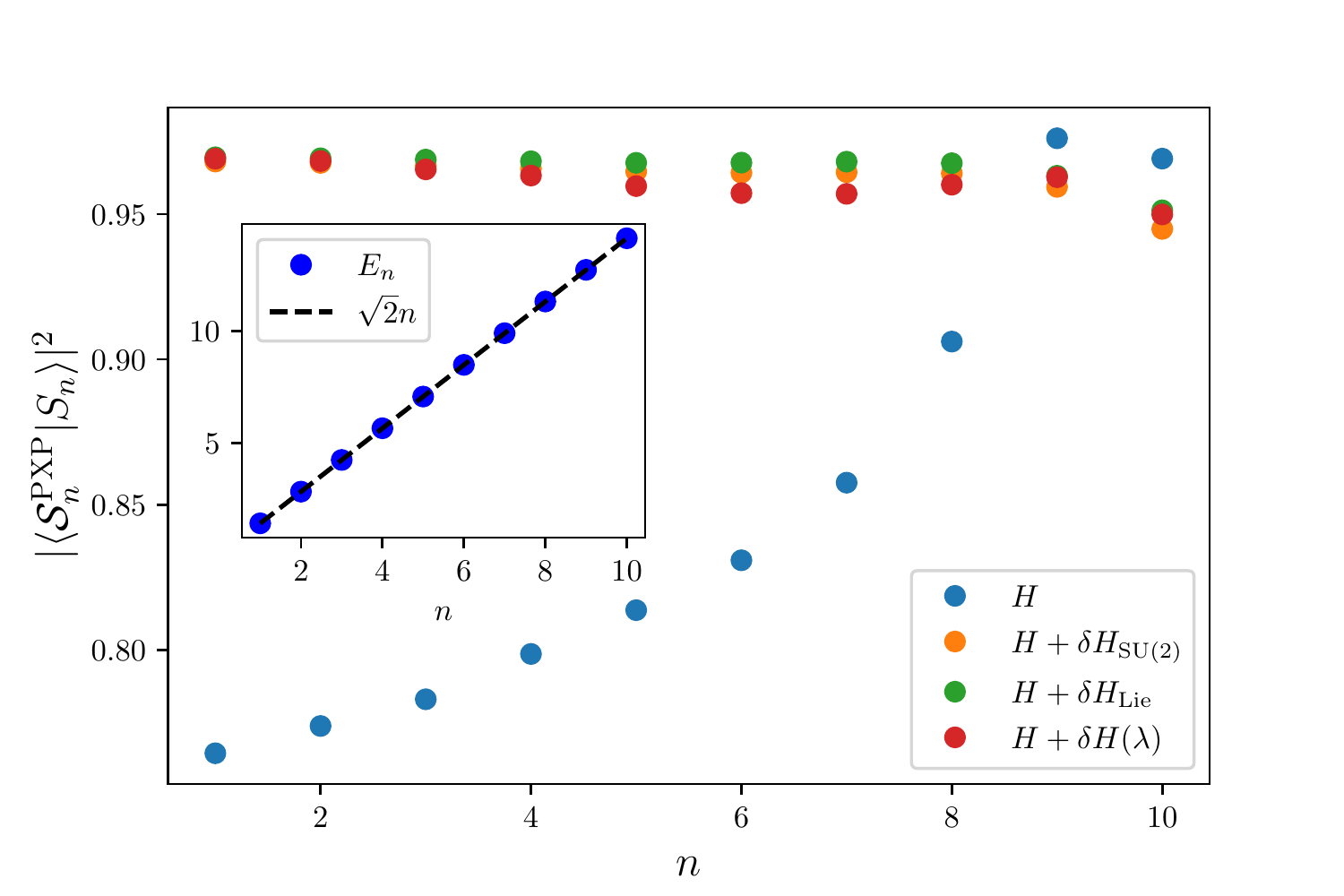}
    \caption{Square of the overlap between the numerically exact scar states and the trial wave functions of Eq.~(\ref{eq:trial state}), i.e., $|\innerproduct*{\mc{S}^{\mr{PXP}}_n}{\mc{S}_n}|^2$ for $N_b=10$ block spins. The numerically exact scar states are obtained by diagonalizing the unperturbed $H$ (blue), $H+\delta H_{\mr{SU}(2)}$ (orange), $H+\delta H_{\mr{Lie}}$ (green), and $H+\delta H(\lambda\approx0.23)$ (red), respectively. Here $\delta H_{\mr{SU}(2)}$ is the perturbation discussed in Ref.~\cite{nearly_su2}. 
    The insert shows the spectrum $\{E_n\}$ of the scar states $\ket*{\mc{S}_n^{\mr{PXP}}}$ in the perturbed PXP model $H+\delta H_{\mr{Lie}}$. ($n=0$ is not included because the level $E=0$ is extensively degenerate, making the choice of $\ket*{\mc{S}_0^{\mr{PXP}}}$ non-unique.).}
    \label{fig:<Sn|Sn> squared}
\end{figure}
\begin{figure}
    \centering
    \includegraphics[width=.48\textwidth]{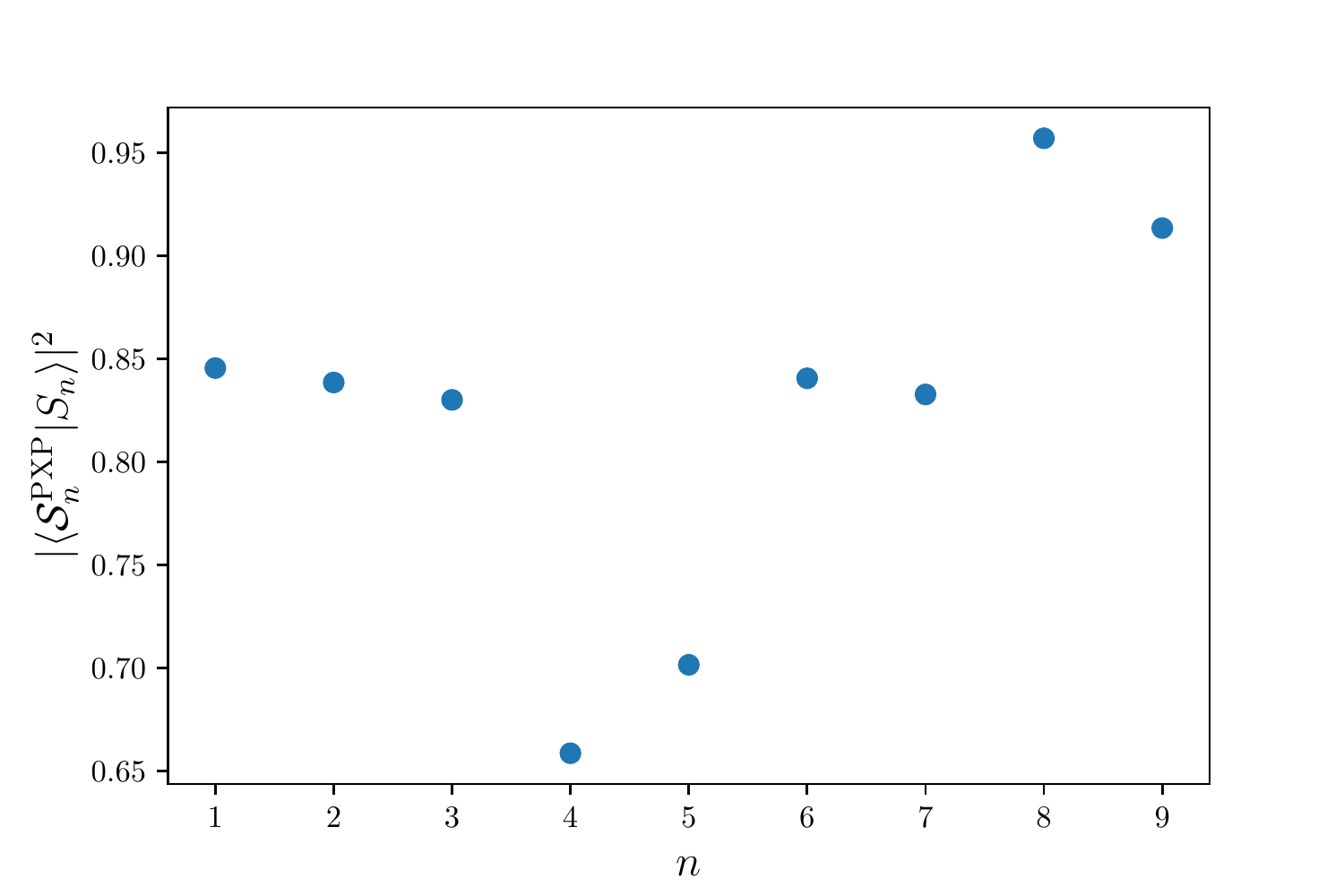}
    \caption{{Square of the overlap $|\innerproduct*{\mc{S}_n^{\mr{PXP}}}{S_n}|^2$ between the exact scar states $\ket*{\mc{S}_n^{\mr{PXP}}}$ and the trial wave functions $\ket*{S_n}$ defined on the honeycomb lattice {of finite size $N_{1} = N_2= 3$ in the direction of the two basis vectors.}}}
    \label{fig:overlap_honeycomb}
\end{figure}

Eq.~(\ref{eq:Neel}) implies a large overlap between $\ket*{\mb{Z}_2}$ and the trial scar states $\ket*{S_n}$. This  suggests they may be good approximations for the exact scar states $\ket*{S_n^{\mr{PXP}}}$, which, in the case of the 1D chain, are indeed known to have large overlap with $\ket*{\mb{Z}_2}$ . This is indeed confirmed by the overlaps we computed between exact and trial scar states, as shown in Fig.~\ref{fig:<Sn|Sn> squared} {for the 1D chain with $N_b=10$ and Fig.~\ref{fig:overlap_honeycomb} for the honeycomb lattice with $N_b=9$ ($N_1=N_2=3$)}. 

\subsection{Energy spacing of scar states}

We now use the states Eq.~(\ref{eq:trial state}) to estimate the energy spacing between neighboring scar states. 
{In general one finds}
\begin{equation}\label{eq:mechanism}
    \begin{split}
        {H\ket*{S_n}}&{=\sqrt{2}n\ket*{S_n}+P_{\mr{Ryd}}H_1\ket*{\wtil{S}_n}}\\
        &{=\left(\sqrt{2}n+\Delta E_n\right)\ket*{S_n}+P_{\mr{Ryd}}\ket*{\delta\wtil{S}_n},}
    \end{split}
\end{equation}
where $\ket*{\delta\wtil{S}_n}$ is the  component of $H_1\ket*{\wtil{S}_n}$ orthogonal to $\ket*{\wtil{S}_n}$. Note that $P_{\mr{Ryd}}H_2\ket*{\wtil{S}_n}$ always vanishes since $P_{\mr{Ryd}}H_2=0$. $\Delta E_n$ is the energy correction arising from $H_1$. 

We estimate these energy corrections below for the 1D chain and the honeycomb lattice.

\subsubsection{{One-dimensional chain}}
As we derive in {Appendix~\ref{app:energy tail}, for the 1D chain,} one finds in the tail of the spectrum, i.e., for $n=N_b,N_b-1$,
\begin{equation}
    \begin{split}
        H{\ket*{S_{N_b}}}&=
        \sqrt{2} {N_b}\left(1-\frac{1}{8}\right){\ket*{S_{N_b}}}+P_{\mr{Ryd}}{\ket*{\delta\wtil{S}_{N_b}}},
    \end{split}
\end{equation}
with a remainder satisfying $\innerproduct*{\delta\wtil{S}_{N_b}}{\wtil{S}_{N_b}}=0$. The leading term is simply the Zeeman eigenvalue, which is reduced by a correction due to $H_1$. For the neighboring scar state one finds
\begin{equation}
    H{\ket*{S_{N_b-1}}}=\left(\frac{7}{8}\sqrt{2}{N_b}-\frac{3}{4}\sqrt{2}\right){\ket{S_{N_b-1}}}+P_{\mr{Ryd}}{\ket*{\delta\wtil{S}_{N_b-1}}},
\end{equation}
with ${\innerproduct*{\delta\wtil{S}_{N_b-1}}{\wtil{S}_{N_b-1}}}=0$.
The thus estimated energies ${E_{N_b}}$ and ${E_{N_b-1}}$ are within 3 percent of the numerically determined values for ${N_b}=|\Lambda|/2=10$ (see Appendix.~\ref{app:energy tail}). The energy spacing between these approximate scar states is estimated as $\Delta E_{\rm tail}=E_{N_b}-E_{N_b-1}\approx(3/4)\sqrt{2}\approx1.06$, which is indeed close to the empirical spacing of scar state energies in the tail of the spectrum. 

{In the middle of the spectrum, i.e., for $n=\order{1}$, the calculation of the energy correction is more involved, but one can obtain an estimate of its asymptotic behavior in the thermodynamic limit ($N_b\rightarrow\infty$) as $\Delta E_{\rm center} = E_1=15/16\sqrt{2}\approx{1.3258},$ very close to the empirical value of $\Omega_{\mr{PXP}}\approx1.33$ (see Appendix.~\ref{app:energy general}).}

We note that a similarly good estimate of this spacing can be obtained from an alternative approximate representation of the scar states, based on matrix product states. These states ${\ket*{\mr{MPS}_n}}$ are obtained by acting with $J^+$ on the exact zero energy state $\ket*{\Gamma}$ from Ref.~\cite{Lin2019mps}, instead of using our approximate trial state $\ket*{\wtil{S}_0}$), and projecting with $P_{\rm Ryd}$ {(see Appendix.~\ref{app:MPS ansatz})}. These states indeed have high overlap, both  with our trial wavefunction $\ket*{{S}_1}$ and with the exact scar state $\ket*{\mc{S}_1^{\mr{PXP}}}$.

For small $n$, the $\ket*{\mr{MPS}_n}$ might be  a slightly better approximation of the scar states of the 1D chain. However, this only holds for the unperturbed PXP model on the 1D chain. For other lattices no similar tensor product states  have been found. Moreover, even for the chain our trial wavefunctions $\ket*{S_n}$ constitute a superior  ansatz, once scar-enhancing perturbations are added to $H_{\mr{PXP}}$.
Indeed, we will see in Sec.~\ref{sec:perturbation} that various perturbations that were discussed in the literature are such that they (partially) cancel the remainder term $H_1$, thus suppressing $\ket*{\delta\wtil{S}_{n}}$ and making the ansatz $\ket*{S_n}$ an even better approximation.
In this sense the $\ket*{S_n}$ capture the essence of the QMBS in the PXP model better than the matrix product states.

\subsubsection{{Honeycomb lattice}}
{For the honeycomb lattice, the energy correction due to $H_1$ differs from the 1D chain. Here we just evaluate it for the tail of the spectrum where one finds}
\begin{equation}\label{eq:estimate honeycomb |SN>}
    {H\ket*{S_{N_b}}=\sqrt{2}N_b\left(1-\frac{7}{32}\right)\ket*{S_{N_b}}+P_{\mr{Ryd}}\ket*{\delta\wtil{S}_{N_b}},}
\end{equation}
{with a remainder satisfying again $\innerproduct*{\delta\wtil{S}_{N_b}}{\wtil{S}_{N_b}}=0$,
while for the neighboring scar state one has}
\begin{equation}\label{eq:estimate honeycomb |SN-1>}
    {H\ket*{S_{N_b-1}}=\left(\frac{25}{32}\sqrt{2}N_b-\frac{15}{32}\sqrt{2}\right)\ket*{S_{N_b-1}}+P_{\mr{Ryd}}\ket*{\delta\wtil{S}_{N_b-1}}}
\end{equation}
suggesting  $\Delta E_{\rm tail}\approx \frac{15}{32}\sqrt{2}$.
Also here the resulting energy estimates $E_{N_b}$ and $E_{N_b-1}$ are within 3 percent of those obtained numerically for $N_b=9$.

The above shows that the ansatz wavefunctions $\ket*{S_n}$ approximately satisfy the stationary Schr\"odinger equation provided that the orthogonal component of $P_{\mr{Ryd}}H_1\ket*{\wtil{S}_n}$ to $\ket*{S_n}$ is small. Moreover the tower of states is nearly equidistant, and the resulting dynamics quasi-periodic for relatively long times, if the $\Delta E_n$ weakly depend on $n$. Both tendencies are favored by lattices with a small coordination number, which controls the number of terms in $H_1$ and thus the importance of deviations from the Zeeman term.

\subsection{{Relations with previous approaches}}
The  oscillations observed in dynamics starting from the N\'eel state have often been interpreted as either  a precession of a large pseudo-spin~\cite{nearly_su2}, or oscillations of a magnon condensate~\cite{Iadecola2019pimagnon}. However, neither of these approaches allows one to obtain the scar states in explicit form nor to predict physical observables such as the energy spacing. Our construction (Eq.~\eqref{eq:trial state} or Eq.~\eqref{eq:|Sn> inv}) fills this gap, and at the same time elucidates explicitly how those pictures are actually realized in Hilbert space.

{On the one hand, Eq.~\eqref{eq:trial state} implies that the scar states are essentially the states of maximal total spin (built from the elementary $S=1$ blocks, not simply the original $S=1/2$ spins), projected onto the constrained subspace. Thus, the oscillations can indeed be interpreted as a large spin ($=$ the maximal total spin) precessing due to the leading Zeeman term (see Fig.~\ref{fig:illustration}). We recall, however, that this picture still refers to a specific dimerization  defining the elementary block spins, even though upon projection our approximate scar wavefunctions are independent of that choice.}

{On the other hand, from the dimerization-invariant expression Eq.~\eqref{eq:|Sn> inv} we can view the scar states as quasi-particle excitations carrying a certain definite lattice momentum (e.g., $k=\pi$ for the 1D chain). Furthermore, one finds that the N\'eel state can be expressed as a coherent state of excitations, namely,}
\begin{equation}
    \begin{split}
        {\ket*{\mb{Z}_2}\propto P_{\mr{Ryd}}U_\varphi\exp\left[-\sum_{i\in\Lambda_\alpha}\sigma_i^-+\sum_{i'\in\Lambda_\beta}\sigma^-_{i'}\right]\bigotimes_{j\in\Lambda}\ket*{\uparrow}_j,}
    \end{split}
\end{equation}
{which one may interpret as a condensate of  excitations.}

Ref.~\cite{Lin2019mps} proposed to understand the scar states as excitations on top of a matrix product state (MPS). This approach is indeed consistent with 
ours, given that our trial wavefunctions $\ket*{S_{\pm N_b}}$ can be written as MPS's (which happen to be idential to the ground state of the Lesanovsky model~\cite{Lesanovsky2012} with the parameter $z=\mp \sqrt{2}$). 

This discussion shows that two apparently contrasting scenarii for the oscillations from the N\'eel state are in fact equivalent: The oscillations can be thought of the precession of a large spin if we express $\ket*{S_n}$ in the  space of $S=1$ block spins, while they can also be seen as the dynamics of a magnon condensate once $\ket*{S_n}$ is written in a dimer-covering invariant manner. We emphasize that it is the structure of the Hamiltonian (Eq.~\eqref{eq:decomposition H}) and its consequence ( Eq.~\eqref{eq:mechanism}) that is essential to understand the tower of scar states in the PXP model. In the following sections Sec.~\ref{sec:perturbation} and \ref{sec:exact perturbation}, we will show that this structure can also be viewed as a generalized projector-embedding (Eq.~\eqref{eq:generalized pe}).

%% file: perturbation.tex
\label{sec:perturbation}
Certain modifications of the PXP Hamiltonian, $H\to H+\delta H$, have been known to stabilize the long-lived  oscillations in dynamics starting from the N\'eel state, {particularly in the 1D chain}; e.g. perturbations of the form $\delta H_d=\lambda_d\sum_{i\in\Lambda_{\mr{B}}}P_{i-1}X_iP_{i+1}(Z_{i-d}+Z_{i+d})$ with $Z\coloneqq\dyad*{\uparrow}-\dyad*{\downarrow}$ and $2\leq d\leq N$ in Ref.~\cite{nearly_su2}, or terms generated by the commutation of parts of the Hamiltonian, which are regarded as ladder operators in an $SU(2)$ algebra, cf.~Ref.~\cite{Kieran2020}.
Below we will use the same notations for the exact scar states as for those of the unperturbed Hamiltonian.
Upon adding the scar-enhancing perturbation $\delta H_{\mr{SU}(2)}$ from Ref.~\cite{nearly_su2} or $\delta H_{\mr{Lie}}$ from Ref.~\cite{Kieran2020} to the PXP Hamiltonian  of a  1D chain, we numerically found very high overlaps $|\innerproduct*{S_n}{\mc{S}^{\mr{PXP}}_n}|^2>0.95$ with our trial states in chains of $N_b=10$ blocks, cf. Fig.~\ref{fig:<Sn|Sn> squared}. 
Moreover, for $\delta H_{\mr{Lie}}$ we found $E_n$ to approach the eigenvalues of $H_Z$, namely $n\sqrt{2}$, up to tiny deviations of order $\order{10^{-3}}$ (see the insert in Fig.~\ref{fig:<Sn|Sn> squared}). In contrast, the trial wavefunctions based on MPS, $\ket*{\mr{MPS}_n}$, become less and less accurate approximations (the squared overlap $|\innerproduct*{\mc{S}_n^{\mr{PXP}}}{\mr{MPS}_n}|^2$ drops to less than $40\%$ , see Fig.~\ref{fig:overlap_MPs}). 

A very high overlap and equidistant energies of $H+\delta H_{\mr{Lie}}$ imply that
\begin{equation}\label{eq:(H+dH)|S>}
\nonumber
    \begin{split}
        (H+\delta H_{\mr{Lie}})\ket*{S_n}&=P_{\mr{Ryd}}\left(H_{\mr{Z}}+H_1+H_2+\delta H_{\mr{Lie}}\right)\ket*{\wtil{S}_{n}}\\
        &
        \approx n\sqrt{2}\ket*{S_n},
    \end{split}
\end{equation}
which in turn reveals the important property that $P_{\mr{Ryd}}\left(H+\delta H_{\mr{Lie}}-H_{\mr{Z}}\right)\ket*{\wtil{S}_n} \equiv P_{\mr{Ryd}}H_{\rm ann}'\ket*{\wtil{S}_n}\approx 0$, where
we have defined
\begin{equation}
H_{\rm ann}' = H_1+H_2+\delta H_{\mr{Lie}}.
\end{equation}
The above suggests that the PXP models {on the 1D chain} with improved scar properties were inadvertedly such that their scar space approached the Rydberg projected maximal spin space spanned by the trial states of Eq.~(\ref{eq:trial state}). This in turn implies that the dynamics starting from the N\'eel state is the projection on ${\cal V}_{\mr{Ryd}}$ of a simple precession {of a spin-coherent state} described by the Zeeman term $H_{\mr{Z}}$ in the effective pseudospin $S=1$ space of block spins, cf. Fig.~\ref{fig:illustration}. 

The above structure suggests how the PXP model fits into the universal framework of QMBS outlaid in \blue{Sec.~\ref{sec:phenomenology}}, provided we suitably generalize it to cases where the Hamiltonian commutes with a global projection operator $P_{\mr{frag}}$  that fragments the Hilbert space. In the present case, we are interested in particular in the Rydberg constraint, i.e.,  $P_{\mr{frag}}=P_{\mr{Ryd}}$.
Indeed, if $H$ possesses a generalized decomposition into a simple spectrum-generating part and an annihilator part, $H=H_{\mr{spec}}+H'_{\mr{ann}}$ (which do not need to commute with the fragmentation individually), the scar states 
can be written as projections 
\begin{equation}
    \ket*{\mc{S}_n}=P_{\mr{frag}}\ket*{\wtil{\mc{S}}_n},
\end{equation} 
{\it provided} that  the parent state $\ket*{\wtil{\mc{S}}_n}$ is (i) an eigenstate of $H_{\mr{spec}}$ with energy $E_n$,
\begin{equation}
H_{\mr{spec}}\ket*{\wtil{\mc{S}}_n}=E_n\ket*{\wtil{\mc{S}}_n},
\end{equation} 
and (ii) is annihilated by $P_{\mr{frag}}H'_{\mr{ann}}$, 
\begin{equation}
P_{\mr{frag}}H'_{\mr{ann}}\ket*{\wtil{\mc{S}}_n}=0.    
\end{equation}
If these conditions are met, $\ket*{\mc{S}_n}$ is indeed an eigenstate of $H$ with energy $E_n$. The framework we outlined in the introduction 
can then be understood as the special case where $P_{\mr{frag}}$ is simply the identity operator. 

As we found empirically above, the "Lie algebra-modified" PXP model, $H_{\mr{mod}}=H+\delta H_{\mr{Lie}}$, approximately possesses such a generalized decomposition, with the  spectrum generating Zeeman term $H_{\mr{spec}}=H_{\mr{Z}}$ and a generalized (approximate) annihilator $H'_{\mr{ann}}\equiv H_{\mr{mod}}-H_{\mr{Z}}=H_1+H_2+\delta H_{\mr{Lie}}$. This observation clarifies in what sense the PXP chain can be understood as being close to a model that shares the common general structure of most other known scar-hosting models.


\section{Exact scars in a non-Hermitian PXP extension and connections with Hermitian perturbations}
\label{sec:exact perturbation}
The observation that projections of maximal spin states are close to the actual scar states suggests how one could construct a  modification of the PXP model having the states $\ket{S_n}$ as {\it exact} scar eigenstates. One simply needs to find a (potentially non-Hermitian) perturbation $H\to H +\delta H_{\mr{NH}}$ which satisfies $P_{\mr{Ryd}}(H_1+\delta H_{\mr{NH}})\ket*{\wtil{S}_n}=0$ for $\forall n$, taking into account that $P_{\mr{Ryd}}H_2=0$. 
We recall that $\ket*{\wtil{S}_n}$ is a maximal spin state, and thus is annihilated by the operators ${1-P_{b,b_1,\cdots,b_k}^{S=k+1}}$, which project out the maximal spin sector  {($S=k+1$) of any set of $k+1$} block spins. Below we will focus on sets consisting of $b$ and the $k$ blocks $b_1,...,b_k$ that are connected by an edge to one of the vertices of $b$). We thus may seek $\delta H_{\mr{NH}}$ such that {$(H_1+\delta H_{\mr{NH}})\prod_bP^{S=k+1}_{b,b_1,\cdots,b_k}=0$}.  
We thus construct an operator that neutralizes the action of $H_1$ on the maximal spin states. 
Eq.~\eqref{eq:pxp block-basis} shows that $H_1$ is a sum of operators, each acting on a block $b$ and the $k$ neighbors of one of its vertices. 
We now construct a neutralizing counterterm for each one of these.  This can be achieved by the choice 
\begin{equation}\label{eq:noneq-pert condition}\begin{split}
{\delta H_{\mr{NH}}}&{=-\sum_{b\in\Lambda_{\mr{B}}}\left(\sum_{k=1}^{z_{b_\alpha}-1}(-1)^k\delta h^{k;\alpha}_b+\sum_{k=1}^{z_{b_\beta}-1}(-1)^k\delta h_b^{k;\beta}\right)}\\
{\delta h_b^{k;\alpha}}&{=\sum_{\substack{\{c_i\}_{i=1}^k\subset\Lambda_{\mr{B}}\\(c_i)_\beta\in\partial b_\alpha\setminus\{b_\beta\}}}\dyad*{0,-,\cdots,-}{\chi_{\{c_i\}}^\alpha}_{b,c_1,\cdots,c_k}}\\
{\delta h_b^{k;\beta}}&{=\sum_{\substack{\{d_i\}_{i=1}^k\subset\Lambda_{\mr{B}}\\(d_i)_\alpha\in\partial b_\beta\setminus\{b_\alpha\}}}\dyad*{0,+,\cdots,+}{\chi_{\{d_i\}}^\beta}_{b,d_1,\cdots,d_k},}
\end{split}
\end{equation}
where {$\ket*{\chi_{\{c_j\}}^\alpha}$ and $\ket*{\chi^\beta_{\{d_j\}}}$} are any wavefunctions of the {($k+1$)} blocks, that {satisfy} 
\begin{equation}\label{eq:counter}
    \begin{split}
        {P_{b,c_1,\cdots,c_k}^{S=k+1}\left(\ket*{\chi_{\{c_i\}}^\alpha}-\ket*{+,-,\cdots,-}\right)_{b,c_1,\cdots,c_k}}&{=0},\\
        {P_{b,d_1,\cdots,d_k}^{S=k+1}\left(\ket*{\chi^\beta_{\{d_i\}}}-\ket*{-,+,\cdots,+}\right)_{b,d_1,\cdots,d_k}}&{=0.}
    \end{split}
\end{equation}
{Indeed, the terms $\delta h_b^{k;\alpha}$ ($\delta h_b^{k;\beta}$) compensate the action of $h_b^{k;\alpha}$ ($h_b^{k;\beta}$) in Eq.~\eqref{eq:H1 expanded} when acting on a maxinmal spin state, as is ensured by Eq.~\eqref{eq:counter}.}

In addition, $\delta H_{\mr{NH}}$ should also  commute with $P_{\rm Ryd}$. This constrains the {$\ket*{\chi^{\alpha,\beta}}$} further, requiring {$\dyad*{-}_b\ket*{\chi^\alpha_{\{c_i\}}}=\dyad*{+}_b\ket*{\chi^\beta_{\{d_i\}}}=0$ and $\dyad*{+}_{c_j}\ket*{\chi^\alpha_{\{c_i\}}}=\dyad{-}_{d_j}\ket*{\chi^\beta_{\{d_i\}}}=0$ for $j=1,\cdots,k$.} These conditions  {potentially admit several solutions. Here we consider a simple one,} 
\begin{equation}
    \begin{split}
        {\ket*{\chi^\alpha_{\{c_i\}}}_{b,c_1,\cdots,c_k}}&={\frac{1}{2k}\sum_{j=1}^k\ket*{0,-,\cdots,\underbracket[1pt][1pt]{0}_{c_j},\cdots,-}_{b,c_1,\cdots,c_k}}\\
        {\ket*{\chi^\beta_{\{d_i\}}}_{b,d_1,\cdots,d_k}}&{=\frac{1}{2k}\sum_{j=1}^k\ket*{0,+,\cdots,\underbracket[1pt][1pt]{0}_{b_j},\cdots,+}_{b,d_1,\cdots,d_k},}
    \end{split}
\end{equation}
where the bracket indicates that the configuration of block $c_j(d_j)$ in $\ket*{\chi^\alpha}(\ket*{\chi^\beta})$  is $\ket*{0}$, while all other blocks $c_m(d_m)\,\,(m\neq j)$ are in state $\ket*{\pm}$.
The state $\ket*{\chi^\alpha}(\ket*{\chi^\beta})$ is chosen to have the same total spin projection $S^z=k-1$ $(S^z=-k+1)$, as the states $\ket*{+,-,\cdots,-}$ $(\ket*{-,+,\cdots,+})$ appearing in Eq.~\eqref{eq:counter}.
Moreover, we choose them such that their inner product with the state $\ket*{S=k+1,S^z=k-1}(\ket*{S=k+1,S^z=-k+1})$ be the same as that of the states $\ket*{+,-,\cdots,-}(\ket*{-,+,\cdots,+})$. This then establishes that Eq.~\eqref{eq:counter} holds.

The above construction ensures that the extended non-Hermitian model $H_{\mr{NH}}= H+\delta H_{\mr{NH}} \equiv H_Z+ H_{\rm ann}'$ possesses exact, energy-equidistant scar states $\ket*{S_n}=P_{\mr{Ryd}}\ket*{\wtil{S}_n}$ since (i) $H_{\mr{NH}}$ commutes with $P_{\mr{Ryd}}$, 
(ii) the parent states $\ket*{\wtil{S}_n}$ are eigenstates of $H_Z$ and (iii) they are annihilated by $P_{\mr{Ryd}} H_{\rm ann}'= P_{\mr{Ryd}}(H_1+\delta H_{\mr{NH}})$, as ensured by Eq.~\eqref{eq:counter}. 
{The resulting non-Hermitian modification of the PXP model having an exact, energy-equidistant tower of scar states, can be cast into the generalized projector-embedding form}{: For any state $\ket*{\psi}\in\mb{C}^{3\otimes N_b}$, the perturbed Hamiltonian acts on its Rydberg projected component $P_{\mr{Ryd}}{\ket*{\psi}}$} as 
\begin{equation}\label{eq:generalized pe}\begin{split}
    &{\left(H+\delta H_{\mr{NH}}\right)P_{\mr{Ryd}}}{\ket*{\psi}}\\&{=P_{\mr{Ryd}}\left[H_{\mr{Z}}+\sum_{b\in\Lambda_{\mr{B}}}h_b\left(1-P_b^{S=S_{\mr{max}}}\right)\right]}{\ket{\psi},}
    \end{split}
\end{equation}
{where $1-P_b^{S=S_{\mr{max}}}$ is a local projector annihilating the parent states of the scars.}

{We recall that the scar states $\ket*{S_n}$ are invariant under a change of the dimerization used to construct the parent states. This can be used to symmetrize $\delta H_{\mr{NH}}$ and restore the symmetries that were lost by the choice of dimerization (see Appendix.~\ref{sec:appendix inv perturbation} for a detailed discussion). Below we carry this out  explicitly for the chain.} 

\subsubsection{{One-dimensional chain}}\label{sec:1D perturbation}
{For the 1D chain, there is only $k=1$ block neighboring a given vertex, and the above construction yields {$\ket*{\chi^{\alpha,\beta}}=\ket*{0,0}$}. {From Eq.~\eqref{eq:noneq-pert condition}} we see that $\delta H_{\mr{NH}}$ then takes the form} 
\begin{equation}\begin{split}
    {\delta H_{\mr{NH}}}&{=\frac{1}{2}\sum_{b\in\Lambda_{\mr{B}}}\left(\ket*{+,0}+\ket*{0,-}\right)\bra*{0,0}_{b,b+1}}\\
    &{=\frac{1}{2}\sum_{b\in\Lambda_{\mr{B}}}P_{2b-1}\left(\sigma^+_{2b}P_{2b+1}+P_{2b}\sigma^+_{2b+1}\right)P_{2b+2},}
    \end{split}
\end{equation}
{where $\sigma^+\coloneqq\dyad*{\uparrow}{\downarrow}$. Here we express $\delta H_{\mr{NH}}$ in the basis of the original $S=1/2$ degrees of freedom {whereby the bond $b$ is formed by the sites $(2b-1, 2b)$}. From this we can obtain an associated  translationally invariant perturbation $\delta H^{\mr{inv}}_{\mr{NH}}$ by averaging $\delta H_{\mr{inv}}$ over translations}
\begin{equation}\begin{split}
\label{Hinv}
    {\delta H^{\mr{inv}}_{\mr{NH}}}&{=\frac{1}{2}\left(\delta H_{\mr{NH}}+ T^{-1}\delta H_{\mr{NH}}T\right)}\\
    &{=\frac{1}{4}\sum_{i\in\Lambda}P_{i-1}\sigma^+_iP_{i+1}\left(P_{i-2}+P_{i+2}\right).}
    \end{split}
\end{equation}
{which still satisfies the requirements (i)-(iii). Moreover,}
{$\ket*{S_n}$ is still an exact eigenstate of $H+\delta H^{\mr{inv}}_{\mr{NH}}$, since $\ket*{S_n}$ is translational invariant.} 

\subsubsection{{Honeycomb lattice}}\label{sec:honeycomb perturbation}
{For the honeycomb lattice, $\delta H_{\mr{NH}}$ becomes}
\begin{equation}
    \begin{split}
        {\delta H_{\mr{NH}}}&{=\frac{1}{2}\sum_{\Vec{R}\in\Lambda_{\mr{B}}}\sum_{i=1,2}\left(\ket*{+,0}+\ket*{0,-}\right)\bra*{0,0}_{\Vec{R},\Vec{R}+\Vec{e}_i}}\\
        &{-\frac{1}{4}\sum_{\Vec{R}\in\Lambda_{\mr{B}}}\ket*{0,-,-}\left(\bra*{0,0,-}+\bra*{0,-,0}\right)_{\Vec{R},\Vec{R}+\Vec{e}_1,\Vec{R}+\Vec{e}_2}}\\
        &{-\frac{1}{4}\sum_{\Vec{R}\in\Lambda_{\mr{B}}}\ket*{+,+,0}\left(\bra*{0,+,0}+\bra*{+,0,0}\right)_{\Vec{R}-\Vec{e}_1,\Vec{R}-\Vec{e}_2,\Vec{R}}.}
    \end{split}
\end{equation}
{Again, we can obtain a rotationally invariant perturbation $\delta H_{\mr{NH}}^{\mr{inv}}$ by averaging $\delta H_{\mr{NH}}$ over rotations (see Appendix.~\ref{sec:appendix inv perturbation}).}

It would obviously be very challenging to engineer an open system with effective non-Hermitian terms Eq.~(\ref{eq:noneq-pert condition}).  However, the main purpose of our  construction is to explicitly show  that the PXP model is in fact close to a model with a strictly local Hamiltonian that preserves the Rydberg constraint and possesses a tower of exact scar states. 
Moreover our explicit example furnishes an illustration of the general algebraic structure underlying scars in a system with a constraint $P_{\mr{Ryd}}$ that fragments Hilbert space.

\subsection{Hermitian scar enhancements of $H_{\rm PXP}$}

A drawback of the above discussed modification is that $H_{\mr{NH}}$ is non-Hermitian. We may, however, take it as a starting point to seek a {\it Hermitian} scar-enhancing perturbation $\delta H$. This will establish a connection to perturbations that have been proposed previously in the literature for the 1D chain.  


Since the non-Hermitian Hamiltonian has real eigenvalues on the scar states, the have vanishing expectation values for its imaginary part.
Thus, an obvious ansatz for a Hermitian perturbation is simply the real part of $\delta H^{\mr{inv}}_{\mr{NH}}$ in Eq.~(\ref{Hinv}),
\begin{equation}\label{eq:dH2new(lambda)}    
    \delta H(\lambda) =\frac{\lambda}{2}(\delta H^{\mr{inv}}_{\mr{NH}}+\delta H^{\mr{inv}\dagger}_{\mr{NH}})
    ={\frac{\lambda}{8}}\sum_{i\in\Lambda}P_{i-1}X_{i}P_{i+1}(P_{i-2}+P_{i+2}).
\end{equation}
where we allow for a coefficient $\lambda\neq 1$ to optimize the resulting Hamiltonian. This is precisely the perturbation studied in Refs.~\cite{nearly_su2, Kieran2020}. 
It remains to optimize $\lambda$ such that the $\ket*{{S}_n}$ become as close as possible to exact eigenstates. To this end we minimize
$\sum_{n=1}^N\norm*{P_{\mr{Ryd}}(H_{1}+\delta H(\lambda))\ket*{\wtil{S}_n}}^2$.
If we neglect the projection by $P_{\mr{Ryd}}$ the minimization can be carried out analytically in the large $N_b$ limit, yielding $\lambda\approx 1.02$. 
Numerical minimization including the projector instead yields the slightly smaller value $\lambda^*\approx
0.93$ (for $N_b=10$), 
which is close to the values obtained with different optimization criteria in Refs~\cite{nearly_su2,Kieran2020}.
With the thus optimized Hamiltonian $H+\delta H(\lambda^*)$ the overlap between the exact and our approximate scar wavefunctions is very high: for $N_b=10$ we find $|\innerproduct*{\mc{S}^{\mr{PXP}}_n}{S_n}|^2\gtrsim 0.95$, cf. Fig.~\ref{fig:<Sn|Sn> squared}. 

The above again confirms the hidden structure of the scar states in both the PXP model and its scar-enhanced cousins: They are projected maximal spin states, as pictorially illustrated in Fig.~\ref{fig:illustration}. Thereby, the large spin $S_{\rm tot}=N_b$ is composed of $N_b$  block spins $S=1$ formed by two elementary $S=1/2$. The N\'eel state has $S^z_{\rm tot}=N_b$ and thus corresponds to the large spin lying essentially in the equatorial plane with respect to the Zeeman field $\sqrt{2}$ along $e_x$, having approximately $S^x_{\rm tot}\approx 0$. 
Upon initialization in the Néel state or elsewhere close to the scar subspace, the dynamics of the system is explicitly seen to be the precession of the large pseudo-spin around its $x$-axis, whose motion is projected down from the $3N$-dimensional $S=1$ space onto the smaller constrained space ${\cal V}_{\rm Ryd}$, cf.~Fig.~\ref{fig:illustration}. Up to small corrections the precession frequency is set by the coefficient $\sqrt{2}$ of the Zeeman term in the $S=1$ space.

%% file: summary_outlook.tex
The present study elucidates the structure of the anomalous eigenstates that violate the ETH in Rydberg {arrays defined on a wide class of lattices}. The original PXP model, and even more so its various scar-enhanced modifications, exhibit scar states having remarkably high overlap with the analytically given projected maximal spin states, $\ket*{S_n}$ of Eq.~\eqref{eq:trial state}. 

We further showed that the PXP model is proximate to scar-enhanced modifications which fit the universal algebraic structure governing essentially all known models hosting exact towers of scars. The usually encountered decomposition into a simple spectral (Zeeman) term and a sum of local annihilators required, however, a generalization to the case of a fragmented Hilbert space, present in the form of the Rydberg constraint in the PXP model. {This generalized structure is present independently of the lattice, as long as it is bipartite and the two sublattices have equal cardinality. We expect our ansatz for the scar wavefunctions to be close to exact scar states especially in cases where the site coordination number is low. For such cases, we can  rationalize the emergence of scar states in the PXP model. 

By leveraging the discovered structure underlying the scars, we have constructed an explicit non-Hermitian perturbation which preserves the Rydberg constraint and possesses the wavefunctions $\ket*{S_n}$ as {\em exact} scar states. Interestingly, the Hermitian part of this systematically constructed perturbation coincides with modifications that were previously studied in the literature on a more empirical basis.} 

The dynamics resulting in these systems is that of a precessing macro-pseudospin whose motion is projected onto the manifold of states obeying the Rydberg constraint. While the picture of a big spin has been suggested in previous works, we provide here an explicit construction of this spin and the associated scar wavefunctions from the elementary degrees of freedom.

Our construction makes use of an explicit dimerization of the bipartite lattice to define a spin space that extends the Rydberg-constrained manifold and breaks certain lattice symmetries. Interestingly, however, the trial scar states we obtain upon projection to the Rydberg manifold, do not depend on the choice of the dimerization, unlike their pre-images. This raises the interesting open question as to whether there might be a unique, dimerization-independent  way to define a parent spin space whose large-spin states project to the scar states.

We demonstrated the dimerization independence by casting the wavefunctions into an explicit form showing that they consist in specific superpositions of maximal spin states on the two sublattices. It will be interesting to explore this form further and establish connections with approaches that consider bipartite structures with very large connectivity~\cite{PhysRevLett.128.090606}.


In the literature, a variety of Hermitian modifications of $H_{\rm PXP}$ and iterative schemes to construct quasi-local models with exact scar states have been discussed. Even though they have often succeeded in greatly enhancing the periodic motion, it is not yet clear whether and why these schemes really converge or whether they are asymptotic in nature. It would be interesting to re-analyze those schemes within the framework and the general structure underlying scar-hosting models we have uncovered here. 

Likewise it will be interesting to apply the insights of this work to {systems with inequivalent sublattices. In the approach presented here, the formation of block spins on dimers was central. This required the two sublattices to have equal cardinality ($|\Lambda_\alpha|=|\Lambda_\beta|$). However, empirically, scar states were also found  numerically in systems with two inequivalent sublattices ($|\Lambda_\alpha|\not=|\Lambda_\beta|$), such as on the decorated honeycomb lattice. Indeed it has been shown that various two-dimensional lattices of Rydberg atoms exhibit similar quantum revivals of an initial  density-wave state, which can be further stabilized by Floquet engineering~\cite{Bluvsteineabg2530}.
} {In those cases it might be interesting to explore generalizations of the wavefunction ansatz Eq.~\eqref{eq:|Sn> inv}, which does not  rely on a dimerization.
}

%% file: appendix_new.tex
\onecolumngrid
\section{Proof of the dimerization invariance of $\ket*{S_n}$ on bipartite lattices}\label{sec:invariance}
Here we show the dimerization invariance of our trial wavefunctions $\ket*{S_n}$,
\begin{equation}\label{eq:SM |Sn>}
    \ket*{S_n}=P_{\mr{Ryd}}\left(\what{J}^-\right)^{N-n}\bigotimes_{b\in\Lambda_{\mr{B}}}\ket*{\what{+}}_b,
\end{equation}
where $\what{J}^\pm=\mp i\sum_{b\in\Lambda_{\mr{B}}}(S_b^y\pm iS_b^z)$ is the collective ladder operator defined in the main text.  

To show the above, we will transform the action of $J^-$ on the block spin into a form that acts individually on the $S=1/2$ degrees of freedom, cf. Eq.~(\ref{eq:from x basis to varphi compact}) below. From this we will infer that for all $k$ the action of $(J^-)^k$  can be written as a product of dimerization invariant operators within the constrained subspace, which proves the dimerization invariance of $\ket*{S_n}$.

One can express the basis $\ket*{\what{\pm}}$ and $\ket*{\what{0}}$ in terms of wavefunctions of the two constituting $S=1/2$ spins as follows: 
\begin{equation}\label{eq:from x basis to S=1/2}
    \begin{split}
        \ket*{\what{+}}_b&=\frac{1}{2}\ket*{+}_b+\frac{1}{\sqrt{2}}\ket*{0}_b+\frac{1}{2}\ket*{-}_b=\frac{1}{2}\ket*{\downarrow\uparrow}_{b_\alpha,b_\beta}+\frac{1}{\sqrt{2}}\ket*{\downarrow\downarrow}_{b_\alpha,b_\beta}+\frac{1}{2}\ket*{\uparrow\downarrow}_{b_\alpha,b_\beta}\\
        \ket*{\what{0}}_b&=\frac{1}{\sqrt{2}}\ket*{+}_b-\frac{1}{\sqrt{2}}\ket*{-}_b=\frac{1}{\sqrt{2}}\ket*{\downarrow\uparrow}_{b_\alpha,b_\beta}-\frac{1}{\sqrt{2}}\ket*{\uparrow\downarrow}_{b_\alpha,b_\beta}\\
        \ket*{\what{-}}_b&=\frac{1}{2}\ket*{+}_b-\frac{1}{\sqrt{2}}\ket*{0}_b+\frac{1}{2}\ket*{-}_b=\frac{1}{2}\ket*{\downarrow\uparrow}_{b_\alpha,b_\beta}-\frac{1}{\sqrt{2}}\ket*{\downarrow\downarrow}_{b_\alpha,b_\beta}+\frac{1}{2}\ket*{\uparrow\downarrow}_{b_\alpha,b_\beta}.
    \end{split}
\end{equation}
The lowering operator $J_b^-\coloneqq\sqrt{2}(\dyad*{\what{0}}{\what{+}}+\dyad*{\what{-}}{\what{0}})_b$, which satisfies $\sum_{b\in\Lambda_{\mr{B}}}J_b^-=J^-$, 
can be used to write $\ket*{\what{0}}_b=\frac{1}{\sqrt{2}}J^-_b\ket*{\what{+}}_b$ and $\ket*{\what{-}}_b=\left(\frac{1}{\sqrt{2}}J^-_b\right)^2\ket*{\what{+}}_b$.

To re-write Eq.~\eqref{eq:from x basis to S=1/2}, and thus the action of $J_b^-$, in a translationally invariant manner, we introduce a new non-orthogonal basis of $S=1/2$ states, and an associated basis transformation $U^\varphi$:
\begin{equation}\begin{split}
    \ket*{\varphi}_i&\coloneqq\cos\varphi\ket*{\uparrow}_i+\sin\varphi\ket*{\downarrow}_i,\,\,\ket*{-\varphi}_i\coloneqq\cos\varphi\ket*{\uparrow}_i-\sin\varphi\ket*{\downarrow}_i,\,\,\varphi\coloneqq\arctan\sqrt{2},\\
    U_i^\varphi&\coloneqq\dyad*{\varphi}{\uparrow}_i+\dyad*{-\varphi}{\downarrow}_i.
    \end{split}
\end{equation}
With this basis, Eq.~\eqref{eq:from x basis to S=1/2} can be written as follows:
\begin{equation}\label{eq:from x basis to varphi}\begin{split}
    \ket*{\what{+}}_b&=\frac{3}{2\sqrt{2}}(1-\dyad*{\uparrow\uparrow})_{b_\alpha, b_\beta}\ket*{\varphi}_{b_\alpha}\ket*{\varphi}_{b_\beta}\\&=\frac{3}{2\sqrt{2}}(1-\dyad*{\uparrow\uparrow})_{b_\alpha, b_\beta}U^\varphi_{b_\alpha}U^\varphi_{b_\beta}\ket*{\uparrow\uparrow}_{b_\alpha, b_\beta}\\
    \ket*{\what{0}}_b&=\frac{3}{2\sqrt{2}}(1-\dyad*{\uparrow\uparrow})_{b_\alpha, b_\beta}\frac{1}{\sqrt{2}}\left(\ket*{\varphi}_{b_\alpha}\ket*{-\varphi}_{b_\beta}-\ket*{-\varphi}_{b_\alpha}\ket*{\varphi}_{b_\beta}\right)\\
    &=\frac{3}{2\sqrt{2}}(1-\dyad*{\uparrow\uparrow})_{b_\alpha, b_\beta}U_{b_\alpha}^\varphi U^\varphi_{b_\beta}\frac{1}{\sqrt{2}}\left(\ket*{\uparrow\downarrow}-\ket*{\downarrow\uparrow}\right)_{b_\alpha, b_\beta}\\
    &=\frac{3}{2\sqrt{2}}(1-\dyad*{\uparrow\uparrow})_{b_\alpha, b_\beta}U_{b_\alpha}^\varphi U^\varphi_{b_\beta}\frac{1}{\sqrt{2}}\left(-\sigma_{b_\alpha}^-+\sigma_{b_\beta}^-\right)\ket*{\uparrow\uparrow}_{b_\alpha, b_\beta}\\
    \ket*{\what{-}}_b&=\frac{3}{2\sqrt{2}}(1-\dyad*{\uparrow\uparrow})_{b_\alpha, b_\beta}\left(-\ket*{-\varphi}_{b_\alpha}\ket*{-\varphi}_{b_\beta}\right)\\&=\frac{3}{2\sqrt{2}}(1-\dyad*{\uparrow\uparrow})_{b_\alpha, b_\beta}U^\varphi_{b_\alpha}U^\varphi_{b_\beta}(-\ket*{\downarrow\downarrow}_{b_\alpha, b_\beta})\\
    &=\frac{3}{2\sqrt{2}}(1-\dyad*{\uparrow\uparrow})_{b_\alpha, b_\beta}U_{b_\alpha}^\varphi U^\varphi_{b_\beta}\left(\frac{1}{\sqrt{2}}(-\sigma_{b_\alpha}^-+\sigma_{b_\beta}^-)\right)^2\ket*{\uparrow\uparrow}_{b_\alpha, b_\beta}.
    \end{split}
\end{equation}
These results can be summarized in the following compact form: 
\begin{equation}
\label{eq:from x basis to varphi compact}
(J^-)^k \ket*{\what{+}}_b = \frac{3}{2\sqrt{2}}(1-\dyad*{\uparrow\uparrow})_{b_\alpha, b_\beta}U_{b_\alpha}^\varphi U^\varphi_{b_\beta}\left(-\sigma_{b_\alpha}^-+\sigma_{b_\beta}^-\right)^k\ket*{\uparrow\uparrow}_{b_\alpha, b_\beta},
\end{equation}
which holds for all $k\geq 0$

Using Eq.~\eqref{eq:from x basis to varphi compact} in 
Eq.~\eqref{eq:SM |Sn>}, one immediately obtains
\begin{equation}\label{eq:manifest inv Sn}
    \ket*{S_n}=\left(\frac{3}{2\sqrt{2}}\right)^NP_{\mr{Ryd}}\prod_{i\in\Lambda}U_i^\varphi\left(\sum_{j\in\Lambda_\alpha}\sigma_j^-{-\sum_{j'\in\Lambda_\beta}\sigma^-_{j'}}\right)^{N-n}\bigotimes_{k\in\Lambda}\ket*{\uparrow}_k,
\end{equation}
which  is manifestly independent of the choice of dimers. 

For the Rydberg chain, let $T$ be the translation operator.  The above explicit form of $\ket*{S_n}$ shows that
\begin{equation}\label{eq:main claim}
    T\ket*{S_n}=(-1)^{N-n}\ket*{S_n},
\end{equation}
expressing the fact that the trial scar states carry total lattice momentum $(N-n)\pi$ (modulo $2\pi$). 

\section{{Restoring lattice symmetries of non-Hermitian perturbations}}
\label{sec:appendix inv perturbation}
{In this section, we show that the non-Hermitian perturbations constructed in Sec.~\ref{sec:perturbation} can be modified so as to restore symmetries of the original Hamiltonian. We start from the following almost trivial observation: if $\ket*{S_n}$ is an eigenstate of $H+\delta H_{\mr{NH}}$ with eigenvalue $E_n$, 
and if $\ket*{S_n}$ is invariant under $g$, i.e., $O_g\ket*{S_n}=\alpha_g\ket*{S_n}$ with a simple phase factor $\alpha_g$, then $\ket*{S_n}$ is also an eigenstate of {the symmetrized Hamiltonian} $H+O_g\delta H_{\mr{NH}}O_g^{-1}$. This observation  implies that by averaging $\delta H_{\mr{NH}}$ over the symmetry group,}
\begin{equation}\label{eq:invariant perturbation}
    \delta H^{\mr{inv}}_{\mr{NH}}=\frac{1}{|G|}\sum_{g\in G}O_g\delta H_{\mr{NH}}O_g^{-1}.
\end{equation}
{we obtain a symmetry preserving perturbation $\delta H_{\mr{NH}}^{\mr{inv}}$ with the same exact scar states.}

\subsection{{One-dimensional chain}}
{As discussed in Sec.~\ref{sec:1D perturbation}, $\delta H_{\mr{NH}}$ for the 1D chain is}
\begin{equation}
    {\delta H_{\mr{NH}}=\frac{1}{2}\sum_{b\in\Lambda_{\mr{B}}}P_{2b-1}\left(\sigma^+_{2b}P_{2b+1}+P_{2b}\sigma^+_{2b+1}\right)P_{2b+2}.}
\end{equation}
{The original PXP model on the 1D chain is invariant with respect to translation $T$ by a single $S=1/2$ site, i.e., $[H,T]=0$. The modified $S=1$ model obtained after dimerization is instead symmetric only with respect to translation by two $S=1/2$ sites. From Eq.~\eqref{eq:manifest inv Sn}, however, we see that its scar state $\ket*{S_n}$ is nevertheless invariant under translation, i.e., $T\ket*{S_n}=(-1)^{N-n}\ket*{S_n}$. This indicates that we may average the Hamiltonian over all translations, or equivalently, just  over the two elements $\{id, T\}$ where $id$ is the identity operator. Applying this result to Eq.~\eqref{eq:invariant perturbation}, we obtain}
\begin{equation}
    {\delta H_{\mr{NH}}^{\mr{inv}}=\frac{1}{4}\sum_{i\in\Lambda}P_{i-1}\sigma_i^+P_{i+1}\left(P_{i-2}+P_{i+2}\right).}
\end{equation}

\subsection{{Honeycomb lattice}}
\begin{figure}
    \centering
    \includegraphics[width=.8\textwidth]{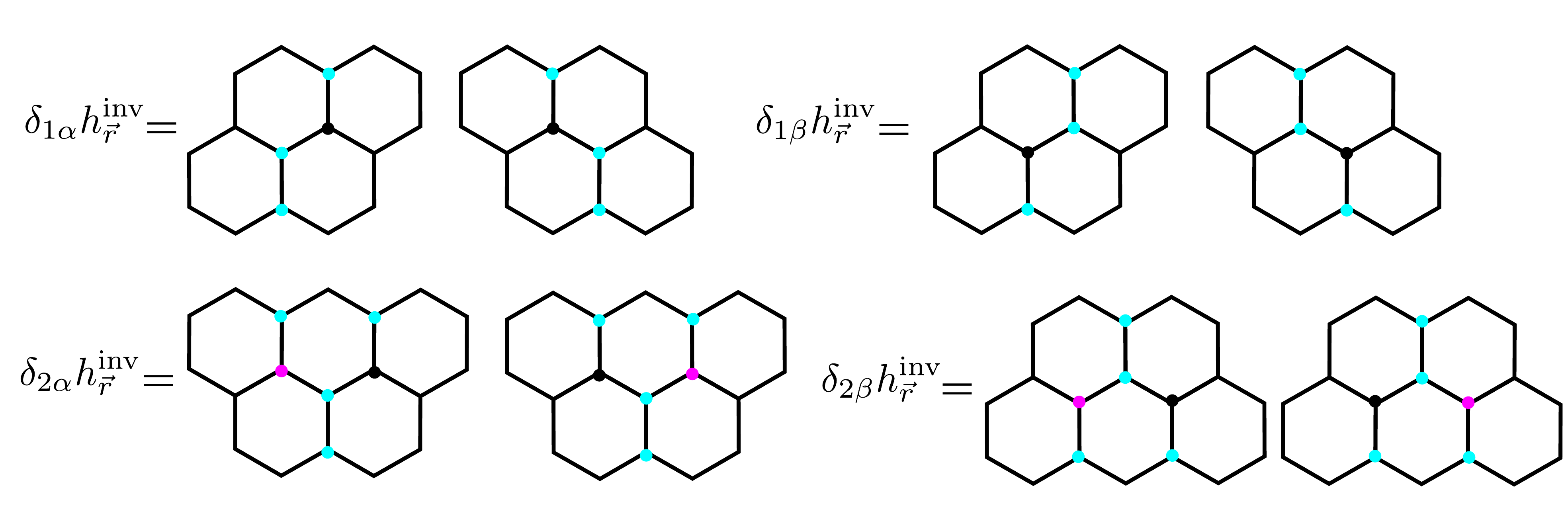}
    \caption{A graphical representation of rotationally invariant perturbations $\delta H_{\mr{NH}}^{\mr{inv}}$. The black, blue and red dots correspond to the operators $\sigma^+$, $P$, and $1-P$, respectively.}
    \label{fig:honeycomb_perturbation}
\end{figure}
{The non-Hermitian perturbation $\delta H_{\mr{NH}}$ for the honeycomb lattice in Sec.~\ref{sec:honeycomb perturbation} can be written as}
\begin{equation}
    \begin{split}
        {\delta H_{\mr{NH}}}&{=\frac{1}{2}\sum_{\Vec{R}\in\Lambda_{\mr{B}}}\sum_{i=1,2}\left(\ket*{+,0}+\ket*{0,-}\right)\bra*{0,0}_{\Vec{R},\Vec{R}+\Vec{e}_i}-\frac{1}{4}\sum_{\Vec{R}\in\Lambda_{\mr{B}}}\ket*{0,-,-}\left(\bra*{0,0,-}+\bra*{0,-,0}\right)_{\Vec{R},\Vec{R}+\Vec{e}_1,\Vec{R}+\Vec{e}_2}}\\
        &{-\frac{1}{4}\sum_{\Vec{R}\in\Lambda_{\mr{B}}}\ket*{+,+,0}\left(\bra*{0,+,0}+\bra*{+,0,0}\right)_{\Vec{R}-\Vec{e}_1,\Vec{R}-\Vec{e}_2,\Vec{R}}}\\
        &{\equiv\frac{1}{2}\sum_{\Vec{R}\in\Lambda_{\mr{B}}}\sum_{i=1,2}\delta_1h_{\Vec{R},\Vec{R}+\Vec{e}_i}-\frac{1}{4}\sum_{\Vec{R}\in\Lambda_{\mr{B}}}\delta_{2\alpha}h_{\Vec{R},\Vec{R}+\Vec{e}_1,\Vec{R}+\Vec{e}_2}-\frac{1}{4}\sum_{\Vec{R}\in\Lambda_{\mr{B}}}\delta_{2\beta}h_{\Vec{R}-\Vec{e}_1,\Vec{R}-\Vec{e}_2,\Vec{R}}}.
    \end{split}
\end{equation}
{As discussed in the main text, the PXP model on the honeycomb lattice possesses a site-centered rotation symmetry. We denote the corresponding rotation operator by $O_g$, and consider $G=\{id, g, g^2\}$. Applying this to Eq.~\eqref{eq:invariant perturbation}, we obtain the rotationally symmetric perturbation}
\begin{equation}
    {\delta H^{\mr{inv}}_{\mr{NH}}=\frac{1}{6}\sum_{\Vec{r}\in\Lambda_\alpha}\delta_{1\alpha}h^{\mr{inv}}_{\Vec{r}}+\frac{1}{6}\sum_{\Vec{r}\in\Lambda_\beta}\delta_{1\beta}h^{\mr{inv}}_{\Vec{r}}-\frac{1}{12}\sum_{\Vec{r}\in\Lambda_\alpha}\delta_{2\alpha}h^{\mr{inv}}_{\Vec{r}}-\frac{1}{12}\sum_{\vec{r}\in\Lambda_\beta}\delta_{2\beta}h^{\mr{inv}}_{\vec{r}},}
\end{equation}
{where}
\begin{equation}\begin{split}
    {\delta_{1\alpha}h_{\Vec{r}}^{\mr{inv}}}&{=\sum_{g\in G}O_g\left(P_{\Vec{r}+\Vec{e}_y}\sigma^+_{\Vec{r}}P_{\Vec{r}-\Vec{e}_1+\Vec{e}_y}P_{\Vec{r}-\Vec{e}_1}+P_{\vec{r}+\vec{e}_y}\sigma_{\vec{r}}^+P_{\vec{r}-\vec{e}_2+\vec{e}_y}P_{\vec{r}-\vec{e}_2}\right)O_g^{-1}}\\
    {\delta_{1\beta}h_{\Vec{r}}^{\mr{inv}}}&{=\sum_{g\in G}O_g\left(P_{\Vec{r}-\Vec{e}_y}\sigma^+_{\Vec{r}}P_{\Vec{r}+\Vec{e}_1-\Vec{e}_y}P_{\Vec{r}+\Vec{e}_1}+P_{\vec{r}-\vec{e}_y}\sigma^+_{\vec{r}}P_{\vec{r}+\vec{e}_2-\vec{e}_y}P_{\vec{r}+\vec{e}_2}\right)O_g^{-1}}\\ 
    {\delta_{2\alpha}h^{\mr{inv}}_{\Vec{r}}}&{=\sum_{g\in G}O_g\left(P_{\Vec{r}-\Vec{e}_1+\Vec{e}_y}P_{\Vec{r}-\Vec{e}_1}\sigma_{\Vec{r}}^+P_{\Vec{r}+\Vec{e}_y}\left(1-P_{\Vec{r}-\Vec{e}_1+\Vec{e}_2}\right)P_{\Vec{r}-\Vec{e}_1+\Vec{e}_2+\Vec{e}_y}+P_{\Vec{r}-\Vec{e}_1+\Vec{e}_y}P_{\Vec{r}-\Vec{e}_1}\left(1-P_{\Vec{r}}\right)P_{\Vec{r}+\Vec{e}_y}\sigma^+_{\Vec{r}-\Vec{e}_1+\Vec{e}_2}P_{\Vec{r}-\Vec{e}_1+\Vec{e}_2+\Vec{e}_y}\right)O_g^{-1}}\\
    {\delta_{2\beta}h^{\mr{inv}}_{\vec{r}}}&{=\sum_{g\in G}O_g\left(P_{\Vec{r}-\Vec{e}_y}\sigma_{\Vec{r}}^+\left(1-P_{\Vec{r}+\Vec{e}_1-\Vec{e}_2}\right)P_{\Vec{r}+\vec{e}_1-\vec{e}_2-\vec{e}_y}P_{\vec{r}+\vec{e}_1+\vec{e}_y}+P_{\Vec{r}-\Vec{e}_y}\left(1-P_{\Vec{r}}\right)\sigma^+_{\Vec{r}+\Vec{e}_1-\Vec{e}_2}P_{\Vec{r}+\vec{e}_1-\vec{e}_2-\vec{e}_y}P_{\vec{r}+\vec{e}_1+\vec{e}_y}\right)O_g^{-1}.}
    \end{split}
\end{equation}
{Fig.~\ref{fig:honeycomb_perturbation} illustrates the above operators graphically.}

\section{Alternative derivation of known exact eigenstates of the PXP model}\label{app:ML state}
There are several exact eigenstates known for the PXP model~\cite{Lin2019mps}: one specific zero-energy eigenstate for PBC, and four eigenstates for open boundary condition (OBC). Their wavefunctions were originally expressd as matrix product states (MPS), and it was pointed out that a basis transformation connects the zero energy eigenstate to the ground state of the AKLT model. Later, Ref.~\cite{Shiraishi_2019} gave an elegant proof of that state being a zero energy eigenstate, by showing that the basis transformation yields {the projector-embedding} form $H=\sum_i\mc{P}_ih_i\mc{P}_i+H'$ where $\mc{P}_i$ is a local projection operator that annihilates the AKLT ground state, and so does $H'$. 

In this section, we give an alternative  proof of these exact eigenstates based on the block spin representation introduced in the main text.
\subsection{Enlarging the Hilbert space}

{In order to show that the states introduced in Ref.~\cite{Lin2019mps} are eigenstates, ``fractional" spins representation of the $S=1$ element play a vital role.} Here we introduce a map which identifies a subspace of the $S=1/2$ model with the $S=1$ block spin~\cite{Tasaki2020}:
\begin{equation}
    A_b\coloneqq\ket*{+}_b\bra*{\uparrow\uparrow}_{(b,\mr{L}),(b,\mr{R})}+\ket*{0}_b\frac{1}{\sqrt{2}}\left(\bra*{\uparrow\downarrow}+\bra*{\downarrow\uparrow}\right)_{(b,\mr{L}),(b,\mr{R})}+\ket*{-}_b\bra*{\downarrow\downarrow}_{(b,\mr{L}),(b,\mr{R})},
\end{equation}
where $(b,\mr{L/R})\in\Lambda_{\mr{B}}^{\mr{frac}}\coloneqq\Lambda_{\mr{B}}\times\{\mr{L},\mr{R}\}$ labels the lattice sites for ``fractional" $S=1/2$ elements. 
We also define the natural tensor product of this map over blocks as $\msf{A}\coloneqq\prod_{b\in\Lambda_{\mr{B}}}A_b:\mb{C}^{2\otimes2\otimes N}\rightarrow\mb{C}^{3\otimes N}$.  

We can fractionalize the spin operator as well. For an arbitrary operator $O:\mb{C}^{3\otimes N}\rightarrow\mb{C}^{3\otimes N}$ of $S=1$ units, there exists an operator $O_{\mr{frac}}:\mb{C}^{2\otimes2\otimes N}\rightarrow\mb{C}^{2\otimes2\otimes N}$ such that $O\msf{A}=\msf{A}O_{\mr{frac}}$. Obviously the choice of $O_{\mr{frac}}$ is not unique. For example, both operators $\sqrt{2}\dyad*{\uparrow\uparrow}{\uparrow\downarrow}_{(b,\mr{L}),(b,\mr{R})}$ and $\sqrt{2}\dyad*{\uparrow\uparrow}{\downarrow\uparrow}_{(b,\mr{L}),(b,\mr{R})}$ correspond to $\dyad*{+}{0}$. However, there is a natural choice for $O_{\mr{frac}}$: we first note that for any operator-valued function $F$, it holds that
\begin{equation}\label{eq:operator identity}
    F(\{\bm{S}_b\}_{b\in\Lambda_{\mr{B}}})\msf{A}=\msf{A}F\left(\left\{\bm{S}_{(b,\mr{L})}+\bm{S}_{(b,\mr{R})}\right\}_{b\in\Lambda_{\mr{B}}}\right),
\end{equation}
where $\bm{S}_b$ is a spin operator with $S=1$, i.e., $\bm{S}_b:\mb{C}^3\rightarrow\mb{C}^3$, and $\bm{S}_{(b,\mr{L/R})}$ is a spin operator with $S=1/2$, i.e., $\bm{S}_{(b,\mr{L/R})}:\mb{C}^2\rightarrow\mb{C}^2$. 
\subsection{Defining exact eigenstates}
\begin{figure}
    \centering
    \includegraphics[width=.65\textwidth]{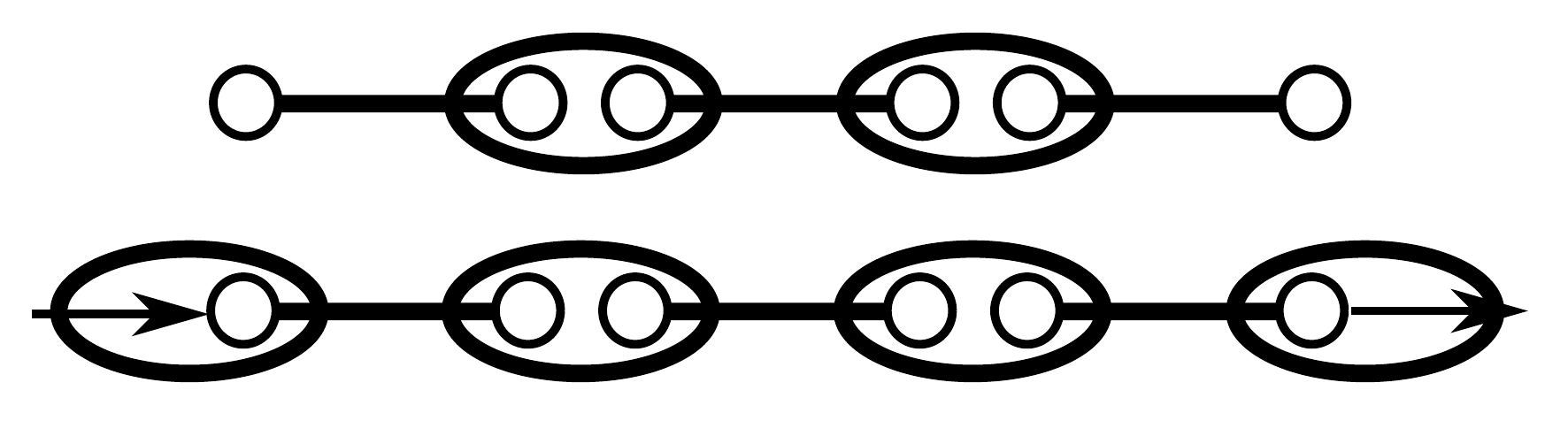}
    \caption{Pictorial representations of $\ket*{\Gamma}$ and $\ket*{\Gamma^{\rightarrow\rightarrow}}$. A line corresponds to the triplet state $(\ket*{\uparrow\uparrow}-\ket*{\downarrow\downarrow})/\sqrt{2}$, while  an oval indicates the symmetrization $\msf{A}$.}
    \label{fig:VBS}
\end{figure}

We will show that the following states $\ket*{\Gamma}$ and $\ket*{\Gamma^{\tau\tau'}}$ are eigenstates of the unperturbed PXP model with PBC and with OBC, respectively:
\begin{equation}\label{eq:|Gamma> in SM}
    \begin{split}
        \ket*{\Gamma}&=\msf{A}\bigotimes_{b\in\Lambda_{\mr{B}}}\frac{1}{\sqrt{2}}\left(\ket*{\uparrow\uparrow}-\ket*{\downarrow\downarrow}\right)_{(b,\mr{R}),(b+1,\mr{L})}\\
        \ket*{\Gamma^{\tau\tau'}}&=\msf{A}\bigotimes_{b\in\Lambda_{\mr{B}}\setminus\{N\}}\frac{1}{\sqrt{2}}\left(\ket*{\uparrow\uparrow}-\ket*{\downarrow\downarrow}\right)_{(b,\mr{R}),(b+1,\mr{L})}\otimes\ket*{\tau}_{(1,\mr{L})}\ket*{\tau'}_{(N,\mr{R})},
    \end{split}
\end{equation}
where $\ket*{\tau}$ and $\ket*{\tau'}$ are either $\ket*{\rightarrow}\coloneqq(\ket*{\uparrow}+\ket*{\downarrow})/\sqrt{2}$ or $\ket*{\leftarrow}\coloneqq(\ket*{\uparrow}-\ket*{\downarrow})/\sqrt{2}$, namely eigenstates of $X(\equiv\dyad*{\uparrow}{\downarrow}+\dyad*{\downarrow}{\uparrow})$. 
Pictorial representations of these states are shown in Fig.~\ref{fig:VBS}. Note the great similarity of the diagrams in Fig.~\ref{fig:VBS} with depictions of the AKLT ground state. 

We first show that these states belong to $\mc{V}_{\mr{Ryd}}$. To do so, we consider the following 2-block states:
\begin{equation}
    \ket*{\gamma^{\sigma\sigma'}_{b,b+1}}\coloneqq \ket{\sigma}_{(b,\mr{L})}\frac{1}{\sqrt{2}}\left(\ket{\uparrow\uparrow}-\ket{\downarrow\downarrow}\right)_{(b,\mr{R}),(b+1,\mr{L})}\ket*{\sigma'}_{b+1,\mr{R}},
\end{equation}
where $\sigma$ and $\sigma'$ are either $\uparrow$ or $\downarrow$, that is, eigenstates of  $Z$. The states $\ket*{\Gamma}$ and $\ket*{\Gamma^{\tau\tau'}}$ can now be expressed in terms  of these $\ket*{\gamma^{\sigma\sigma'}_{b,b+1}}$ as:
\begin{equation}\label{eq:|Gamma> decomposition}
    \begin{split}
         \ket*{\Gamma}&=\msf{A}\sum_{\sigma,\sigma'=\uparrow,\downarrow}c_{\sigma\sigma'}\ket*{\gamma^{\sigma\sigma'}_{b,b+1}}\otimes\ket*{\Xi_{\sigma\sigma'}}\\
         \ket*{\Gamma^{\tau\tau'}}&=\msf{A}\sum_{\sigma,\sigma'=\uparrow,\downarrow}c_{\sigma\sigma'}\ket*{\gamma^{\sigma\sigma'}_{b,b+1}}\otimes\ket*{\Xi^{\tau\tau'}_{\sigma\sigma'}},
    \end{split}
\end{equation}
where $c_{\sigma\sigma'}$ are coefficients and $\ket*{\Xi_{\sigma\sigma'}}$ and $\ket*{\Xi_{\sigma\sigma'}^{\tau\tau'}}$ are spin-$1/2$ wavefunctions defined on $(\Lambda_{\mr{B}}\setminus\{b,b+1\})\times\{\mr{L},\mr{R}\}$.
A straightforward calculation yields
\begin{equation}
    \begin{split}
        A_b\otimes A_{b+1}\ket*{\gamma_{b,b+1}^{\uparrow\uparrow}}&=\frac{1}{\sqrt{2}}\ket*{+,+}_{b,b+1}-\frac{1}{2\sqrt{2}}\ket*{0,0}_{b,b+1}\\
        A_b\otimes A_{b+1}\ket*{\gamma_{b,b+1}^{\uparrow\downarrow}}&=\frac{1}{2}\ket*{+,0}_{b,b+1}-\frac{1}{2}\ket*{0,-}_{b,b+1}\\
        A_b\otimes A_{b+1}\ket*{\gamma_{b,b+1}^{\downarrow\uparrow}}&=\frac{1}{2}\ket*{0,+}_{b,b+1}-\frac{1}{2}\ket*{-,0}_{b,b+1}\\
        A_b\otimes A_{b+1}\ket*{\gamma_{b,b+1}^{\downarrow\downarrow}}&=\frac{1}{2\sqrt{2}}\ket*{0,0}_{b,b+1}-\frac{1}{\sqrt{2}}\ket*{-,-}_{b,b+1}.
    \end{split}
\end{equation}
Therefore, combined with Eq.~\eqref{eq:|Gamma> decomposition}, one finds 
\begin{equation}\label{eq:local Pryd}
    \begin{split}
        (1-\dyad{+,-})_{b,b+1}\ket*{\Gamma}&=+\ket*{\Gamma}\\
        (1-\dyad{+,-})_{b,b+1}\ket*{\Gamma^{\tau\tau'}}&=+\ket*{\Gamma^{\tau\tau'}}
    \end{split}
\end{equation}
for $\forall b\in\Lambda_{\mr{B}}$. Since $P_{\mr{Ryd}}$ is the product over $b$ of $\left(1-\dyad{+,-}\right)_{b,b+1}$ and 
and local Rydberg constraints ($1-\dyad*{+,-}_{b,b+1}$) commute with each other, we find $P_{\mr{Ryd}}\ket*{\Gamma}=+\ket*{\Gamma}$ and $P_{\mr{Ryd}}\ket*{\Gamma^{\tau\tau'}}=+\ket*{\Gamma^{\tau\tau'}}$, implying $\ket*{\Gamma},\ket*{\Gamma^{\tau\tau'}}\in\mc{V}_{\mr{Ryd}}$.

\subsection{A detailed proof of the  eigenstate property}
As $P_{\mr{Ryd}}\ket*{\Gamma}=+\ket*{\Gamma}$, it holds that
\begin{equation}
    H\ket*{\Gamma}=HP_{\mr{Ryd}}\ket*{\Gamma}=P_{\mr{Ryd}}H\ket*{\Gamma}=P_{\mr{Ryd}}\left(H_{\mr{Z}}+H_1\right)\ket*{\Gamma}.
\end{equation}
Here we use $\left[P_{\mr{Ryd}},H\right]=0$ and $P_{\mr{Ryd}}\ket*{+,-}=0$. Further, Eq.~\eqref{eq:local Pryd} implies that $\dyad*{+,-}_{b,b+1}\ket*{\Gamma}=0$ for $\forall b\in\Lambda_{\mr{B}}$, and thus $H_1 \ket*{\Gamma}=0$.
We therefore obtain $H\ket*{\Gamma}=P_{\mr{Ryd}}H_{\mr{Z}}\ket*{\Gamma}$.
The same relation holds for $\ket*{\Gamma^{\tau\tau'}}$ as well. With the operator identity of Eq.~\eqref{eq:operator identity}, we find 
\begin{equation}\begin{split}
    H_{\mr{Z}}\ket{\Gamma}&=\msf{A}\left(\sqrt{2}\sum_{b\in\Lambda_{\mr{B}}}\left(\frac{1}{2}X_{(b,\mr{L})}+\frac{1}{2}X_{(b,\mr{R})}\right)\right)\bigotimes_{b'\in\Lambda_{\mr{B}}}\frac{1}{\sqrt{2}}\left(\ket{\uparrow\uparrow}-\ket{\downarrow\downarrow}\right)_{(b',\mr{R}),(b'+1,\mr{L})}\\
    &=\msf{A}\left(\sqrt{2}\sum_{b\in\Lambda_{\mr{B}}}\left(\frac{1}{2}X_{(b,\mr{R})}+\frac{1}{2}X_{(b+1,\mr{L})}\right)\right)\bigotimes_{b'\in\Lambda_{\mr{B}}}\frac{1}{\sqrt{2}}\left(\ket{\uparrow\uparrow}-\ket{\downarrow\downarrow}\right)_{(b',\mr{R}),(b'+1,\mr{L})}\\&=0,
    \end{split}
\end{equation}
where the PBC was used in the second line. For $\ket*{\Gamma^{\tau\tau'}}$, we find
\begin{equation}
    \begin{split}
        H_{\mr{Z}}\ket*{\Gamma^{\tau\tau'}}&=\msf{A}\left(\sqrt{2}\sum_{b\in\Lambda_{\mr{B}}}\left(\frac{1}{2}X_{(b,\mr{L})}+\frac{1}{2}X_{(b,\mr{R})}\right)\right)\bigotimes_{b'\in\Lambda_{\mr{B}}\setminus\{N\}}\frac{1}{\sqrt{2}}\left(\ket*{\uparrow\uparrow}-\ket*{\downarrow\downarrow}\right)_{(b',\mr{R}),(b'+1,\mr{L})}\otimes\ket*{\tau}_{(1,\mr{L})}\ket*{\tau'}_{(N,\mr{R})}\\
        &=\msf{A}\frac{1}{\sqrt{2}}\left(X_{(1,\mr{L})}+X_{(N,\mr{R})}\right)\bigotimes_{b'\in\Lambda_{\mr{B}}\setminus\{N\}}\frac{1}{\sqrt{2}}\left(\ket*{\uparrow\uparrow}-\ket*{\downarrow\downarrow}\right)_{(b',\mr{R}),(b'+1,\mr{L})}\otimes\ket*{\tau}_{(1,\mr{L})}\ket*{\tau'}_{(N,\mr{R})}.
    \end{split}
\end{equation}
Since $\ket{\tau}$ and $\ket*{\tau'}$ were chosen as eigenstates of $X$, $\ket*{\Gamma^{\tau\tau'}}$ is seen to be an eigenstate of $H_{\mr{Z}}$.

\subsection{Matrix product state (MPS)  representation}
In Ref.~\cite{Lin2019mps}, $\ket*{\Gamma}$ and $\ket*{\Gamma^{\tau\tau'}}$ were given as MPS. Here we re-derive their MPS representation. 
To do so, we define a bond variable $\rket{\alpha}_b$ 
for $b\in\Lambda_{\mr{B}}$ such that
\begin{equation}
    \rket{0}_b\coloneqq\ket*{\uparrow\uparrow}_{(b-1,\mr{R}),(b,\mr{L})},\,\,
    \rket{1}_b\coloneqq\ket*{\downarrow\downarrow}_{(b-1,\mr{R}),(b,\mr{L})}.
\end{equation}
$\ket*{\Gamma}$ is re-written as
\begin{equation}
    \begin{split}
        \ket*{\Gamma}&=\msf{A}\bigotimes_{b\in\Lambda_{\mr{B}}}\frac{1}{\sqrt{2}}\left(\rket{0}-\rket{1}\right)_b.
    \end{split}
\end{equation}
From this expression, we can find the MPS representation for $\ket*{\Gamma}$: a straightforward calculation yields 
\begin{equation}\label{eq:Ab}\begin{split}
    A_b\frac{1}{\sqrt{2}}\left(\rket{0}-\rket{1}\right)_b\otimes\frac{1}{\sqrt{2}}\left(\rket{0}-\rket{1}\right)_{b+1}&=\msf{A}_{1,1}^+\ket*{\uparrow}_{(b-1,\mr{R})}\ket*{+}_b\ket*{\uparrow}_{(b+1,\mr{L})}+\msf{A}_{1,2}^0\ket*{\uparrow}_{(b-1,\mr{R})}\ket*{0}_b\ket*{\downarrow}_{(b+1,\mr{L})}\\
    &-\msf{A}_{2,1}^0\ket*{\downarrow}_{(b-1,\mr{R})}\ket*{0}_b\ket*{\uparrow}_{(b+1,\mr{L})}-\msf{A}_{2,2}^-\ket*{\downarrow}_{(b-1,\mr{R})}\ket*{-}_b\ket*{\downarrow}_{(b+1,\mr{L})},
    \end{split}
\end{equation}
where 
\begin{equation}
\msf{A}_{1,1}^+\coloneqq\frac{1}{2},\,\,\msf{A}^0_{1,2}\coloneqq-\frac{1}{2\sqrt{2}},\,\,\msf{A}^0_{2,1}\coloneqq\frac{1}{2\sqrt{2}},\,\,\msf{A}^-_{2,2}\coloneqq-\frac{1}{2}.
\end{equation}
We emphasize that $\rket{0}_b$ is different from $\ket*{0}_b$ in Eq.~\eqref{eq:Ab}: $\rket{0}_b$ is a bond variable and $\ket*{0}_b$ is an eigenstate of $S^z_b$.
$\ket*{\Gamma}$ can be written as
\begin{equation}
    \ket*{\Gamma}=\sum_{\{\sigma_i\}_{i=1}^N\in\{\pm,0\}^{\otimes N}}\mr{Tr}\left[\msf{A}^{\sigma_1}\cdots\msf{A}^{\sigma_{N}}\right]\ket*{\sigma_1\cdots\sigma_N},
\end{equation}
where 
\begin{equation}
    \msf{A}^+\coloneqq\frac{1}{2\sqrt{2}}\begin{pmatrix}\sqrt{2} & 0\\0&0\end{pmatrix},\,\,\msf{A}^0\coloneqq\frac{1}{2\sqrt{2}}\begin{pmatrix}0&-1\\1&0\end{pmatrix},\,\,\msf{A}^-\coloneqq\frac{1}{2\sqrt{2}}\begin{pmatrix}0&0\\0&-\sqrt{2}\end{pmatrix}.
\end{equation}
Up to an irrelevant normalization factor, this is the same as the zero energy state given in Ref.~\cite{Lin2019mps}.

A similar calculation leads to the MPS representation of $\ket*{\Gamma^{\tau\tau'}}$:
\begin{equation}
    \begin{split}
        \ket*{\Gamma^{\tau\tau}}=\sum_{\{\sigma_i\}_{i=1}^N\in\{\pm,0\}^{\otimes N}}\left(\msf{v}_\tau^{\sigma_1}\right)^{\mr{T}}\msf{A}^{\sigma_2}\cdots\msf{A}^{\sigma_{N-1}}\msf{w}_{\tau'}^{\sigma_N}\ket*{\sigma_1\cdots\sigma_N},
    \end{split}
\end{equation}
where 
\begin{equation}
    \begin{split}
        \msf{v}_\rightarrow^+&\coloneqq\frac{1}{2}\begin{pmatrix}\sqrt{2}\\0\end{pmatrix},\,\,\msf{v}_\rightarrow^0\coloneqq\frac{1}{2}\begin{pmatrix}1\\-1\end{pmatrix},\,\,\msf{v}_\rightarrow^-\coloneqq\frac{1}{2}\begin{pmatrix}0\\-\sqrt{2}\end{pmatrix},\,\,\msf{v}^+_\leftarrow\coloneqq\frac{1}{2}\begin{pmatrix}\sqrt{2}\\0\end{pmatrix},\,\,\msf{v}^0_\leftarrow\coloneqq\frac{1}{2}\begin{pmatrix}-1\\-1\end{pmatrix},\,\,\msf{v}_\leftarrow^-\coloneqq\frac{1}{2}\begin{pmatrix}0\\\sqrt{2}\end{pmatrix},\\
        \msf{w}_\rightarrow^+&\coloneqq\frac{1}{2\sqrt{2}}\begin{pmatrix}\sqrt{2}\\0\end{pmatrix},\,\,\msf{w}_\rightarrow^0\coloneqq\frac{1}{2\sqrt{2}}\begin{pmatrix}1\\1\end{pmatrix},\,\,\msf{w}_\rightarrow^-\coloneqq\frac{1}{2\sqrt{2}}\begin{pmatrix}0\\\sqrt{2}\end{pmatrix},\,\,\msf{w}_\leftarrow^+\coloneqq\frac{1}{2\sqrt{2}}\begin{pmatrix}\sqrt{2}\\0\end{pmatrix},\,\,\msf{w}_\leftarrow^0\coloneqq\frac{1}{2\sqrt{2}}\begin{pmatrix}-1\\1\end{pmatrix},\,\,\msf{w}_\leftarrow^-\coloneqq\frac{1}{2\sqrt{2}}\begin{pmatrix}0\\-\sqrt{2}\end{pmatrix}.
    \end{split}
\end{equation}
The relations
\begin{equation}\begin{split}
\begin{pmatrix}1&1\end{pmatrix}\msf{A}^\sigma&=\frac{1}{\sqrt{2}}\msf{v}^\sigma_\rightarrow,\,\,\begin{pmatrix}1&-1\end{pmatrix}\msf{A}^\sigma=\frac{1}{\sqrt{2}}\msf{v}^\sigma_\leftarrow,\,\,\msf{A}^\sigma\begin{pmatrix}1\\1\end{pmatrix}=\msf{w}^\sigma_\leftarrow,\,\,\msf{A}^\sigma\begin{pmatrix}1\\-1\end{pmatrix}=\msf{w}^\sigma_\rightarrow,
\end{split}
\end{equation}
finally establish the equivalence of the states $\ket*{\Gamma^{\tau\tau}}$
to the four eigenstates given in Ref.~\cite{Lin2019mps}.
\subsection{Alternative set of trial wave functions}
In the main text, we consider the trial wave functions $\ket*{S_n}$, which are viewed as projections on ${\cal V}_{\rm Ryd}$ of the eigenstates of $H_{\mr{Z}}$ with maximal total pseudospin. However, one can also construct an alternative set of trial wave functions based on the exact reference scar state $\ket*{\Gamma}$. We consider the ansatz
\begin{equation}\label{eq:another Sn}
    \ket*{\mr{MPS}_n}\coloneqq P_{\mr{Ryd}}\left(J^+\right)^n\ket*{\Gamma},
\end{equation}
where $J^\pm$ is defined in 
the main text. In the spin-$1/2$ representation, $\ket*{\mr{MPS}_n}$ is expressed as 
\begin{equation}\label{eq:|MPS>}
    \begin{split}
        \ket*{\mr{MPS}_n}&=\sum_{\{b_k\}_{k=1}^n\subset\Lambda_{\mr{B}}}P_{\mr{Ryd}}\msf{A}\bigotimes_{k=1}^n\ket*{\rightarrow\rightarrow}_{(b_k,\mr{R}),(b_k+1,\mr{L})}\bigotimes_{b'\in\Lambda_{\mr{B}}\setminus\{b_k\}_{k=1}^n}\frac{1}{\sqrt{2}}\left(\ket*{\uparrow\uparrow}-\ket*{\downarrow\downarrow}\right)_{(b',\mr{R}),(b'+1,\mr{L})}.
    \end{split}
\end{equation}

\begin{figure}
    \centering
    \includegraphics[width=.6\textwidth]{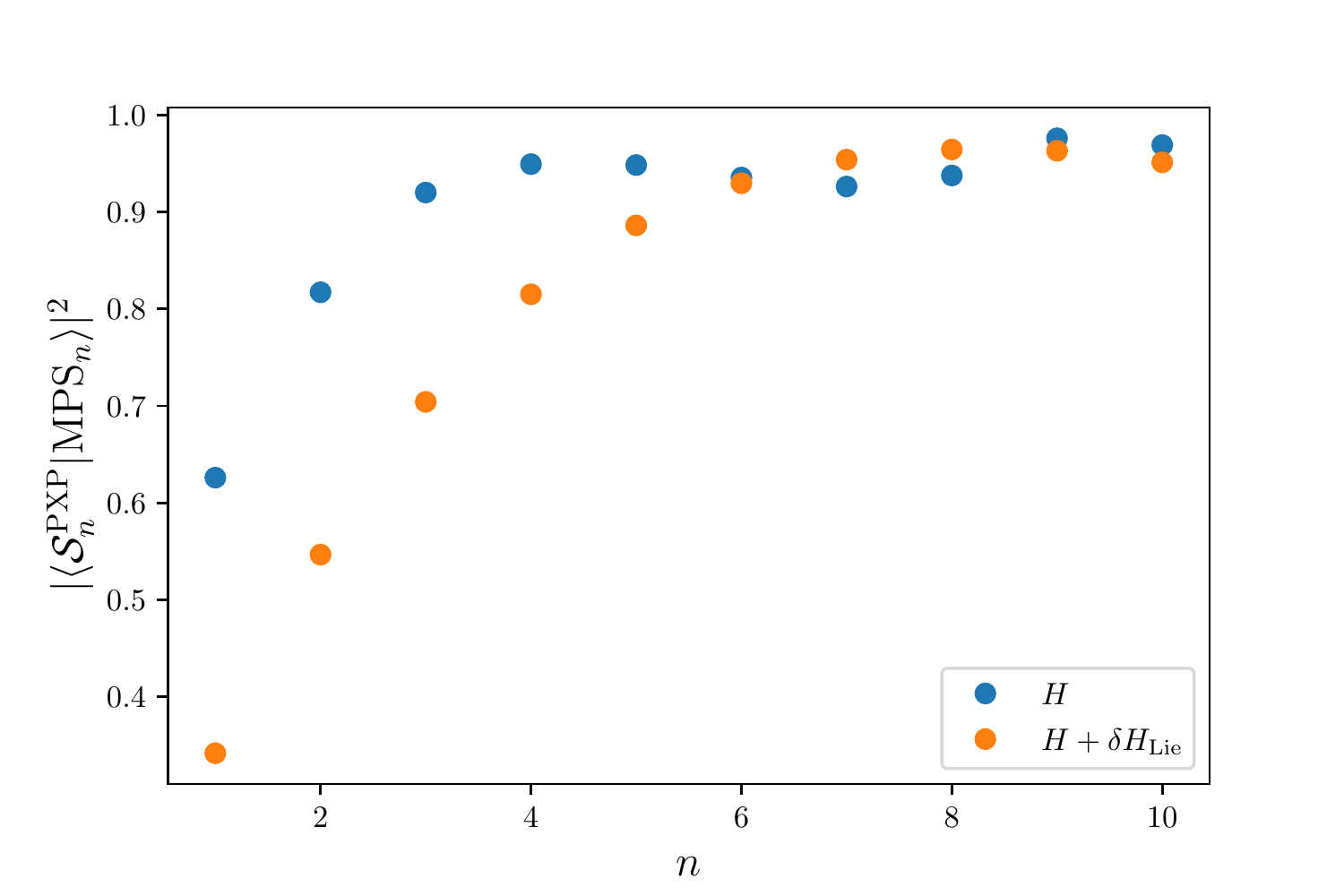}
    \caption{Square of the overlap $|\innerproduct*{\mc{S}_n^{\mr{PXP}}}{\mr{MPS}_n}|^2$ between the exact scar states $\ket*{\mc{S}_n^{\mr{PXP}}}$ and the trial wave functions $\ket*{\mr{MPS}_n}$ from Eq.~(\ref{MPS}).}
    \label{fig:overlap_MPs}
\end{figure}

We numerically find that these states have similarly high overlap with the exact scar states as $\ket*{S_n}$, cf. Fig.~\ref{fig:overlap_MPs}. We note, however, that they become increasingly worse approximations once scar-enhancing perturbations are added to the Hamiltonian.

\section{Energy of the trial scar states}
In this appendix we estimate the energy of the scar states in the tail of the spectrum, i.e., for $\ket*{\mc{S}^{\mr{PXP}}_{N_b}}$ and $\ket*{\mc{S}^{\mr{PXP}}_{N_b-1}}$, as well as in the middle of the spectrum (for $\ket*{\mc{S}^{\mr{PXP}}_1}$), by evaluating the expectation value of $H$ on our approximate trial wavefunctions.
{Here we restrict ourselves to the 1D chain, but the generalization to any lattice, including the honeycomb lattice discussed in the main text, is straightforward.}
\subsection{{Energy estimate using $\ket*{S_n}$}}\label{app:energy general}
{Here we estimate the energy of the scar states using our ansatz $\ket*{S_n}$, based on first-order perturbation theory. As discussed in the main text, one finds}
\begin{equation}
\begin{split}
    H\ket*{S_n}&=P_{\mr{Ryd}}\left(H_{\mr{Z}}+H_1+H_2\right)\ket*{\wtil{S}_n}\\&=P_{\mr{Ryd}}\left(H_{\mr{Z}}+H_1\right)\ket*{\wtil{S}_n}
    \label{eq:perturbation}\\&=n\sqrt{2}\ket*{S_n}+P_{\mr{Ryd}}H_1\ket*{\wtil{S}_n},
    \end{split}
\end{equation}
where $\ket*{\wtil{S}_n}$ is an (unnormalized) eigenstate of the Zeeman term with eigenvalue $n\sqrt{2}$. 
To derive an energy correction, we do not directly compute the expectation value of $H$ in the ansatz state $\ket*{S_n}$. Instead we consider the second line of Eq.~\eqref{eq:perturbation} and observe that $\ket*{\wtil{S}_n}$ is an eigenstate  of the Zeeman term $H_Z$ 
within the spin-1 space. We then 
ask  by how much $H_1$ shifts the corresponding eigenvalue within first porder perturbation theory. This is expected to give a good estimate of the energy shift for $\ket*{S_n}$ as well, but allows us to circumvent the difficulties related to the Rydberg constraint.
  We thus regard $H_{\mr{Z}}$ and $\ket*{\wtil{S}_n}$ as the unperturbed Hamiltonian, and the unperturbed states, respectively, and $H_1$ as a (non-Hermitian) perturbation. The first order energy correction is given as
\begin{equation}\label{eq:perturbation2}
    \begin{split}
        {\Delta E_n=\frac{\expval*{H_1}{\wtil{S}_n}}{\mc{N}_n} =\frac{1}{\mc{N}_n}\sum_b\bra*{\wtil{S}_n}\left(\ket*{+,0}+\ket*{0,-}\right)\bra*{+,-}_{b,b+1}\ket*{\wtil{S}_n},}
    \end{split}
\end{equation}
{where $\mc{N}_n=\innerproduct*{\wtil{S}_n}$ is the norm of the (non-normalized) parent scar wavefunction. 
We now compute the correction from each local term ($(\ket*{+,-}+\ket*{0,-})\bra*{+,-}$). To do so, we first re-write the ansatz $\ket*{\wtil{S}_n}$ for the $n$'th scar state in a suitable way,} 
\begin{equation}\begin{split}
   \ket*{\wtil{S}_n}&=
    \left(J^-\right)^{N-n}\bigotimes_{b\in\Lambda_{\mr{B}}}\ket*{\what{+}}_b,
    \\
    J^\pm&=\sqrt{2}\sum_{b\in\Lambda_{\mr{B}}}\left(\dyad*{\what{\pm}}{\what{0}}+\dyad*{\what{0}}{\what{\mp}}\right)_b.
    \end{split}
\end{equation}
{We split the collective spin-raising (lowering) operator as}
\begin{equation}\begin{split}
    {J^\pm}&{=J_{b,b+1}^\pm+J_{\Lambda_{\mr{B}}\setminus\{b,b+1\}}^\pm},\\
    {J_{b,b+1}^\pm}&{\coloneqq\sqrt{2}\left(\dyad*{\what{\pm}}{\what{0}}+\dyad*{\what{0}}{\what{\mp}}\right)_b+\sqrt{2}\left(\dyad*{\what{\mp}}{\what{0}}+\dyad*{\what{0}}{\what{\mp}}\right)_{b+1}},\\
    {J^\pm_{\Lambda_{\mr{B}}\setminus\{b,b+1\}}}&{\coloneqq\sqrt{2}\sum_{b'\in\Lambda_{\mr{B}}\setminus\{b,b+1\}}\left(\dyad*{\what{\pm}}{\what{0}}+\dyad*{\what{0}}{\what{\mp}}\right)_{b'}.}
    \end{split}
\end{equation}
{Using these operators, we can write $\ket*{\wtil{S}_{N-n}}$ as}
\begin{equation}\label{eq:|Sn> split}\begin{split}
    {\ket*{\wtil{S}_{N-n}}}&{=\sum_{k=0}^{\min\{n,4\}}\binom{n}{k}\left(J^-_{b,b+1}\right)^k\ket*{\what{T}_{2,2}}_{b,b+1}\otimes\left(J^-_{\Lambda_{\mr{B}}\setminus\{b,b+1\}}\right)^{n-k}\ket*{\what{T}_{N-2,N-2}}_{\Lambda_{\mr{B}}\setminus\{b,b+1\}}}\\
    &{=\sum_{m=\max\{2-n,-2\}}^2c_m\ket*{\what{T}_{2,m}}_{b,b+1}\otimes\ket*{\what{T}_{N-2,N-n-m}}_{\Lambda_{\mr{B}}\setminus\{b,b+1\}},}
    \end{split}
\end{equation}
where $c_m$ are numerical constants and $\ket*{\what{T}_{S,M}}$ is a state with total spin $S$ and $S^x=M$. Below we will mostly need the coefficients $c_{\pm2}$ for $n\geq4$, which take the values
\begin{eqnarray}
\label{c2_1}
    c_{2}&=&\prod_{M=N_b-n-1}^{N_b-2}\sqrt{(N_b-2)(N_b-1)-M(M-1)},\\c_{-2}&=&n(n-1)(n-2)(n-3)\prod_{M=N_b-n+3}^{N_b-2}\sqrt{(N_b-2)(N_b-1)-M(M-1)}.\nonumber
\end{eqnarray}
{The matrix element of a local term of $H_1$, which we denote by $\delta e^m_b$, is obtained as}
\begin{equation}
    {\delta e^m_b\coloneqq-\bra*{\what{T}_{2,m}}\left(\ket*{+,0}+\ket*{0,-}\right)\braket*{+,-}{\what{T}_{2,m}}=}\begin{dcases}
    -\frac{\sqrt{2}}{8}&\,\,m=+2,\\
    +\frac{\sqrt{2}}{8}&\,\,m=-2,\\
    0&\,\,\mr{else}.
    \end{dcases}
\end{equation}
{The energy correction from all the sites is thus estimated as,} 
\begin{equation}
\label{eq:energy correction formula}
    {\Delta E_{N_b-n}=\sum_{b\in\Lambda_{\mr{B}}}\sum_{m=-2}^2\delta e^m_b\frac{|c_m|^2}{\mc{N}_{N_b-n}}=-\frac{\sqrt{2}}{8}N_b\frac{|c_2|^2-|c_{-2}|^2}{\mc{N}_{N_b-n}}.}
\end{equation}
{From the elementary theory of angular momentum, the norm $\mc{N}_{N_b-n}$ is easily calculated,}
\begin{equation}
\label{norm}
    {\mc{N}_{N_b-n}=\prod_{M=N_b-n+1}^{N_b}\left(N_b(N_b+1)-M(M-1)\right).}
\end{equation}

\subsubsection{{The tail of the scar spectrum}}\label{app:energy tail}
{The estimate of $E_{N_b}$ and $E_{N_b-1}$ is relatively straightforward. Indeed, for {$n=0$}, one immediately finds $|c_2|^2=\mc{N}_{N_b}$ and $c_{-2}=0$. Therefore the energy correction becomes $\Delta E_{N_b}=-\sqrt{2}/8\times N_b$. Thus, the energy of the state $\ket*{S_n}$ is estimated as $E_{N_b}=7\sqrt{2}N_b/8$. 
For $N_b=10$, our estimate yields $E_{N_b}=70\sqrt{2}/8\approx12.37$, which is within 3 percent of the numerically determined eigenvalue of the exact scar state, $E_{N_b}\approx12.07$. 

For {$n=1$}, a straightforward calculation yields $|c_2|^2/\mc{N}_{N_b-1}=(N_b-2)/N_b$, {and also here $c_{-2}=0$}, which implies $\Delta E_{N_b-1}=-\sqrt{2}/8\times(N_b-2)$. Thus, we obtain the estimate $E_{N_b-1}=(7N_b-6)\sqrt{2}/8$. For $N_b=10$, this yields $11.37$, which is close to the numerical value $E_{N_b-1}\approx11.10$. {The energy spacing in the tails of the spectrum can thus be estimated as $\Delta E_{\mr{tail}}=E_{N_b}-E_{N_b-1}\approx(3/4)\sqrt{2}\approx1.06$, which is within 10 percent of the numerically determined value.} 

\subsubsection{{Center of the scar spectrum}}
{We can also estimate the energy in the middle of the spectrum in the large $N_b$ limit. As an example, we estimate $E_1$. From Eq.~\eqref{c2_1}, $|c_{\pm2}|^2$ is expressed as}
\begin{equation}
\label{c2s}
\begin{split}
    {|c_2|^2}&{=\prod_{M=0}^{N_b-2}{\left((N_b-2)(N_b-1)-M(M-1)\right),\,|c_{-2}|^2=\prod_{M=0}^3(N_b-1-M)^2\prod_{M=4}^{N_b-2}\left((N_b-2)(N_b-1)-M(M-1)\right)}}.
    \end{split}
\end{equation}
{From this  one easily finds {$(|c_2|^2-|c_{-2}|^2)/|c_2|^2= 8/N_b+O(1/N_b^2)$} in the large $N_b$ limit. Using Eqs.~(\ref{norm}) and (\ref{c2s}), taking a logarithm and approximating the summation by an integral, one obtains $|c_2|^2/\mc{N}_1 = 1/16 +O(1/N_b)$.} Substituting these values into Eq.~\eqref{eq:energy correction formula}, the energy correction is estimated in the large $N_b$ limit as
\begin{equation}
    {\Delta E_1\approx-\frac{\sqrt{2}}{8}N_b\frac{8}{N_b}\times\frac{1}{16}=-\frac{\sqrt{2}}{16}.}
\end{equation}
Recalling that $E_0=0$, This yields an estimate of the level spacing in the centrum of the scar spectrum as $\Delta E_{\rm center}= E_1 \approx \sqrt{2}+ \Delta E_1= 15\sqrt{2}/16$.

\subsection{The middle of the spectrum using $\ket*{\mr{MPS}_n}$}\label{app:MPS ansatz}
To estimate the energy of $\ket*{\mc{S}_1^{\mr{PXP}}}$, we use the alternative ansatz $\ket*{\mr{MPS}_1}$ in Eq.~\eqref{eq:|MPS>}. Using the MPS representation, $\ket*{\mr{MPS}_1}$ can be written as
\begin{equation}
\label{MPS}
    \ket*{\mr{MPS}_1}=\sum_{b\in\Lambda_{\mr{B}}}\sum_{\{\sigma_i\}_{i=1}^N\in\{\pm,0\}^{\otimes N}}P_{\mr{Ryd}}\left(\msf{v}_\rightarrow^{\sigma_b}\right)^{\mr{T}}\msf{A}^{\sigma_{[b+1]}}\cdots\msf{A}^{\sigma_{[b+N-2]}}\msf{w}_\rightarrow^{\sigma_{[b+N-1]}}\ket*{\sigma_1\cdots\sigma_N},
\end{equation}
where $[b]$ satisfies $b\equiv[b] (\mr{mod}\,\,N)$ and $1\leq [b]<N$. This is conveniently rewritten by defining the wavefunction  $\ket*{M^{\rightarrow\rightarrow}_{b,b+1}}$  as
\begin{equation}\begin{split}
    \ket*{M^{\rightarrow\rightarrow}_{b,b+1}}&\coloneqq\sum_{\{\sigma_i\}_{i=1}^N\in\{\pm,0\}^{\otimes N}}\left(\msf{v}_\rightarrow^{\sigma_b}\right)^{\mr{T}}\msf{A}^{\sigma_{[b+1]}}\cdots\msf{A}^{\sigma_{[b+N-2]}}\msf{w}_\rightarrow^{\sigma_{[b+N-1]}}\ket*{\sigma_1\cdots\sigma_N}\\
    &=\msf{A}\ket*{\rightarrow\rightarrow}_{(b,\mr{R}),(b+1,\mr{L})}\bigotimes_{b'\in\Lambda_{\mr{B}}\setminus\{b\}}\frac{1}{\sqrt{2}}\left(\ket*{\uparrow\uparrow}-\ket*{\downarrow\downarrow}\right)_{(b',\mr{R}),(b'+1,\mr{L})},
    \end{split}
\end{equation}
so that the unprojected MPS parent state is $\ket*{\wtil{\mr{MPS}}_1}=\sum_{b\in\Lambda_{\mr{B}}}\ket*{M^{\rightarrow\rightarrow}_{b,b+1}}$, and one has  $\ket*{\mr{MPS}_1}=P_{\mr{Ryd}}\ket*{\wtil{\mr{MPS}}_1}$ and .
With this we find
\begin{equation}\label{eq:H|S'1>}\begin{split}
    H\ket*{\mr{MPS}_1}&=\sqrt{2}\ket*{\mr{MPS}_1}-\sum_{b\in\Lambda_{\mr{B}}}P_{\mr{Ryd}}\left(\ket*{+,0}+\ket*{0,-}\right)\bra*{+,-}_{b,b+1}\ket*{M^{\rightarrow\rightarrow}_{b,b+1}}\\
    &=\sqrt{2}\ket*{\mr{MPS}_1}+\frac{1}{4}\sum_{b\in\Lambda_{\mr{B}}}\left(\ket*{+,0}+\ket*{0,-}\right)_{b,b+1}\otimes\ket*{M^{\uparrow\downarrow}_{b-1,b+2}},
    \end{split}
\end{equation}
where
\begin{equation}
    \begin{split}
        \ket*{M_{b-1,b+2}^{\uparrow\downarrow}}\coloneqq\left(\bigotimes_{b'\in\Lambda_{\mr{B}}\setminus\{b,b+1\}}A_{b'}\right)\ket*{\uparrow\downarrow}_{(b-1,\mr{R}),(b+2,\mr{L})}\bigotimes_{b''\in\Lambda_{\mr{B}}\setminus\{b-1,b,b+1\}}\frac{1}{\sqrt{2}}\left(\ket*{\uparrow\uparrow}-\ket*{\downarrow\downarrow}\right)_{(b'',\mr{R}),(b''+1,\mr{L})}.
    \end{split}
\end{equation}
The MPS representation of $\ket*{M^{\uparrow\downarrow}_{b-1,b+2}}$ is as follows:
\begin{equation}
    \ket*{M_{b-1,b+2}^{\uparrow\downarrow}}=\sum_{\{\sigma_i\}_{i\in\Lambda_{B}\setminus\{b,b+1\}}\in\{\pm,0\}^{\otimes N-2}}\left(\msf{v}^{\sigma_{[b+2]}}_\downarrow\right)^{\mr{T}}\msf{A}^{\sigma_{[b+3]}}\cdots\msf{A}^{\sigma_{[b-2]}}\msf{w}_\uparrow^{\sigma_{[b-1]}}\ket*{\sigma_1\cdots\sigma_{N}},
\end{equation}
where $\sigma_b$ and $\sigma_{b+1}$ are excluded from the summation. We have defined the boundary vectors as $\msf{v}_\downarrow^\sigma\coloneqq(\msf{v}^\sigma_\rightarrow-\msf{v}^\sigma_\leftarrow)/\sqrt{2}$ and $\msf{w}^\sigma_\uparrow\coloneqq(\msf{w}^\sigma_\rightarrow+\msf{w}^\sigma_\leftarrow)/\sqrt{2}$. We denote the second term in the second line in Eq.~\eqref{eq:H|S'1>} as $\ket*{\delta\mr{MPS}_1}$, i.e., $H\ket*{\mr{MPS}_1}=\sqrt{2}\ket*{\mr{MPS}_1}+\frac{1}{4}\ket*{\delta\mr{MPS}_1}$. Thus, the energy expectation value of $\ket*{\mr{MPS}_1}$ can be evaluated from 
\begin{equation}\label{eq:renormalized H|S'1>}
    H\ket*{\mr{MPS}_1}=\left(\sqrt{2}+\frac{1}{4}\frac{\innerproduct*{\mr{MPS}_1}{\delta\mr{MPS}_1}}{\innerproduct*{\mr{MPS}_1}{\mr{MPS}_1}}\right)\ket*{\mr{MPS}_1}+\ket*{\mr{MPS}_\perp},
\end{equation}
where $\ket*{\mr{MPS}_\perp}$ satisfies $\innerproduct*{\mr{MPS}_1}{\mr{MPS}_\perp}=0$. Since both $\ket*{\mr{MPS}_1}$ and $\ket*{\delta\mr{MPS}_1}$ can be expressed as MPS, one can calculate their norm and inner product by standard techniques for MPS states. 
For large $N$ one finds,
\begin{equation}
    \innerproduct*{\mr{MPS}_1}\cong\frac{14N}{9}\left(\frac{3}{4}\right)^N,\,\,\innerproduct*{\mr{MPS}_1}{\delta\mr{MPS}_1}\cong-\frac{4\sqrt{2}N}{9}\left(\frac{3}{4}\right)^N.
\end{equation}
Thus, for large $N$, we obtain $\sqrt{2}+\frac{1}{4}\frac{\innerproduct*{\mr{MPS}_1}{\delta\mr{MPS}_1}}{\innerproduct*{\mr{MPS}_1}}\cong\frac{13}{14}\sqrt{2}\cong1.3132$, which is indeed very close to the empirical level spacing ($\Omega_{\rm PXP}\cong1.33$) between scar states close to the center of the spectrum.

To quantify the quality of the approximation of the trial scar state, one can estimate $\norm*{\ket*{\mr{MPS}_\perp}}$ and obtains $\norm*{\ket*{\mr{MPS}_\perp}}/\norm*{\ket*{\mr{MPS}_1}}=\sqrt{151/(3\cdot14^3)}\cong0.1354$ in the thermodynamic limit.  

\section{Estimate of optimal  scar-enhancement}
In the main text we have quoted the optimal coefficient for the perturbation $\delta H(\lambda)$ (Eq.~\eqref{eq:dH(lambda)} or Eq.~\eqref{eq:dH2(lambda)}). Here we provide a detailed derivation.
The actual goal is to minimize $\sum_n\norm*{P_{\mr{Ryd}}(H_1+\delta H(\lambda))\ket*{S_n}}^2$ with respect to $\lambda$, which we carried out numerically for $N=10$ blocks. Here, we instead want to gain more analytical understanding. 
To keep the task tractable we drop the projection on the Rydberg subspace and restrict ourselves to  minimizing $\sum_n\norm*{(H_1+\delta H(\lambda))\ket*{\wtil{S}_n}}^2$, under the assumption that the two optimization problems do not differ too much.

\subsection{The scar states}
The trial wave function $\ket{S_n}$ for the $n$'th scar state is as follows,
\begin{equation}\begin{split}
    \ket{S_n}&=P_{\mr{Ryd}}\ket*{\wtil{S}_n}=P_{\mr{Ryd}}\left(J^-\right)^{N-n}\bigotimes_{b\in\Lambda_{\mr{B}}}\ket*{\what{+}}_b
    \\
    J^\pm&=\sqrt{2}\sum_{b\in\Lambda_{\mr{B}}}\left(\dyad*{\what{\pm}}{\what{0}}+\dyad*{\what{0}}{\what{\mp}}\right)_b.
    \end{split}
\end{equation}
We split the collective spin-raising (lowering) operator as
\begin{equation}\begin{split}
    J^\pm&=J_{b,b+1}^\pm+J_{\Lambda_{\mr{B}}\setminus\{b,b+1\}}^\pm\\
    J_{b,b+1}^\pm&\coloneqq\sqrt{2}\left(\dyad*{\what{\pm}}{\what{0}}+\dyad*{\what{0}}{\what{\mp}}\right)_b+\sqrt{2}\left(\dyad*{\what{\mp}}{\what{0}}+\dyad*{\what{0}}{\what{\mp}}\right)_{b+1}\\
    J^\pm_{\Lambda_{\mr{B}}\setminus\{b,b+1\}}&\coloneqq\sqrt{2}\sum_{b'\in\Lambda_{\mr{B}}\setminus\{b,b+1\}}\left(\dyad*{\what{\pm}}{\what{0}}+\dyad*{\what{0}}{\what{\mp}}\right)_{b'}.
    \end{split}
\end{equation}
Using these operators, we can write $\ket*{\wtil{S}_{N-n}}$ with $n\geq3$ as
\begin{equation}\begin{split}
    \ket*{\wtil{S}_{N-n}}&=\sum_{k=0}^4\binom{n}{k}\left(J^-_{b,b+1}\right)^k\ket*{\what{T}_{2,2}}_{b,b+1}\otimes\left(J^-_{\Lambda_{\mr{B}}\setminus\{b,b+1\}}\right)^{n-k}\ket*{\what{T}_{N-2,N-2}}_{\Lambda_{\mr{B}}\setminus\{b,b+1\}}\\
    &=c\ket*{\what{T}_{2,2}}_{b,b+1}\otimes\ket*{\what{T}_{N-2,N-n-2}}_{\Lambda_{\mr{B}}\setminus\{b,b+1\}}+2c\sqrt{\frac{n}{-n+2N-3}}\ket*{\what{T}_{2,1}}_{b,b+1}\otimes\ket*{\what{T}_{N-2,N-n-1}}_{\Lambda_{\mr{B}}\setminus\{b,b+1\}}\\
    &+c\sqrt{\frac{3n!(-n+2N-4)!}{2(n-2)!(-n+2N-2)!}}\ket*{\what{T}_{2,0}}_{b,b+1}\otimes\ket*{\what{T}_{N-2,N-n}}_{\Lambda_{\mr{B}}\setminus\{b,b+1\}}\\
    &+c\sqrt{\frac{n!(-n+2N-4)!}{6(n-3)!(-n+2N-1)!}}\ket*{\what{T}_{2,-1}}_{b,b+1}\otimes\ket*{\what{T}_{N-2,N-n+1}}_{\Lambda_{\mr{B}}\setminus\{b,b+1\}}\\
    &+c\sqrt{\frac{n!(-n+2N-4)!}{(n-4)!(-n+2N)!}}\ket*{\what{T}_{2,-2}}_{b,b+1}\otimes\ket*{\what{T}_{N-2,N-n+2}}_{\Lambda_{\mr{B}}\setminus\{b,b+1\}},
    \end{split}
\end{equation}
where $c\coloneqq\prod_{M=N-n-1}^{N-2}\sqrt{(N-2)(N-1)-M(M-1)}$. Here, $\ket*{\what{T}_{S,M}}$ is a state with total spin $S$ and $S^x=M$. When $N$ is sufficiently large, one can approximate $\ket*{\wtil{S}_n}$ as
\begin{equation}\label{eq:|Sn> approximate}
    \begin{split}
        \frac{1}{c}\ket*{\wtil{S}_{N-n}}&\cong \ket*{\what{T}_{2,2}}_{b,b+1}\otimes\ket*{\what{T}_{N-2,N-n-2}}_{\Lambda_{\mr{B}}\setminus\{b,b+1\}}+2\sqrt{\frac{s}{2-s}}\ket*{\what{T}_{2,1}}_{b,b+1}\otimes\ket*{\what{T}_{N-2,N-n-1}}_{\Lambda_{\mr{B}}\setminus\{b,b+1\}}\\
        &+\sqrt{\frac{3}{2}}\frac{s}{2-s}\ket*{\what{T}_{2,0}}_{b,b+1}\otimes\ket*{\what{T}_{N-2,N-n}}_{\Lambda_{\mr{B}}\setminus\{b,b+1\}}+\sqrt{\frac{s^3}{6(2-s)^3}}\ket*{\what{T}_{2,-1}}_{b,b+1}\otimes\ket*{\what{T}_{N-2,N-n+1}}_{\Lambda_{\mr{B}}\setminus\{b,b+1\}}\\
        &+\frac{1}{12}\frac{s^2}{(2-s)^2}\ket*{\what{T}_{2,-2}}_{b,b+1}\otimes\ket*{\what{T}_{N-2,N-n+2}}_{\Lambda_{\mr{B}}\setminus\{b,b+1\}},
    \end{split}
\end{equation}
where $s\coloneqq n/N$. Each coefficient is plotted in Fig.~\ref{fig:coeffs}. When $n$ is small, the dominant contribution in $\ket*{\wtil{S}_{N-n}}$ is $\ket*{\what{T}_{2,2}}_{b,b+1}$, but as $n$ increases $\ket*{\what{T}_{2,1}}_{b,b+1}$ and $\ket*{\what{T}_{2,0}}_{b,b+1}$ become dominant. For later arguments, we define the reduced density matrix $\rho_S$ for the incoherent equal weight distribution over the states $\ket*{\wtil{S}_n}$ as 
\begin{equation}
    \rho_S\coloneqq\frac{1}{N}\sum_{n=1}^N\mr{Tr}_{\Lambda_{\mr{B}}\setminus\{b,b+1\}}\frac{1}{\norm*{\ket*{\wtil{S}_n}}^2}\dyad*{\wtil{S}_n},
\end{equation}
where $\mr{Tr}_{\Lambda_{\mr{B}}\setminus\{b,b+1\}}$ is a partial trace on $\Lambda_{\mr{B}}\setminus\{b,b+1\}$. We can approximately obtain $\rho_S$ using Eq.~\eqref{eq:|Sn> approximate} as
\begin{equation}
    \begin{split}
        \rho_S&\cong\frac{1}{N}\sum_{s=1/N}^{1}\frac{1}{Z(s)}\left(P^{(2,2)}_{b,b+1}+\frac{4s}{2-s}P^{(2,1)}_{b,b+1}+\frac{3s^2}{2(2-s)^2}P^{(2,0)}_{b,b+1}+\frac{s^3}{6(2-s)^3}P^{(2,-1)}_{b,b+1}+\frac{s^4}{144(2-s)^4}P^{(2,-2)}_{b,b+1}\right)\\
        Z(s)&\coloneqq1+\frac{4s}{2-s}+\frac{3s^2}{2(2-s)^2}+\frac{s^3}{6(2-s)^3}+\frac{s^4}{144(2-s)^4},
    \end{split}
\end{equation}
where $P_{b,b+1}^{(S,M)}\coloneqq\dyad*{\what{T}_{S,M}}_{b,b+1}$. We can approximate $\rho_S$ further by replacing $N^{-1}\sum_{s=1/N}^1\rightarrow\int_0^1ds$.
\begin{figure}
    \centering
    \includegraphics[width=.5\textwidth]{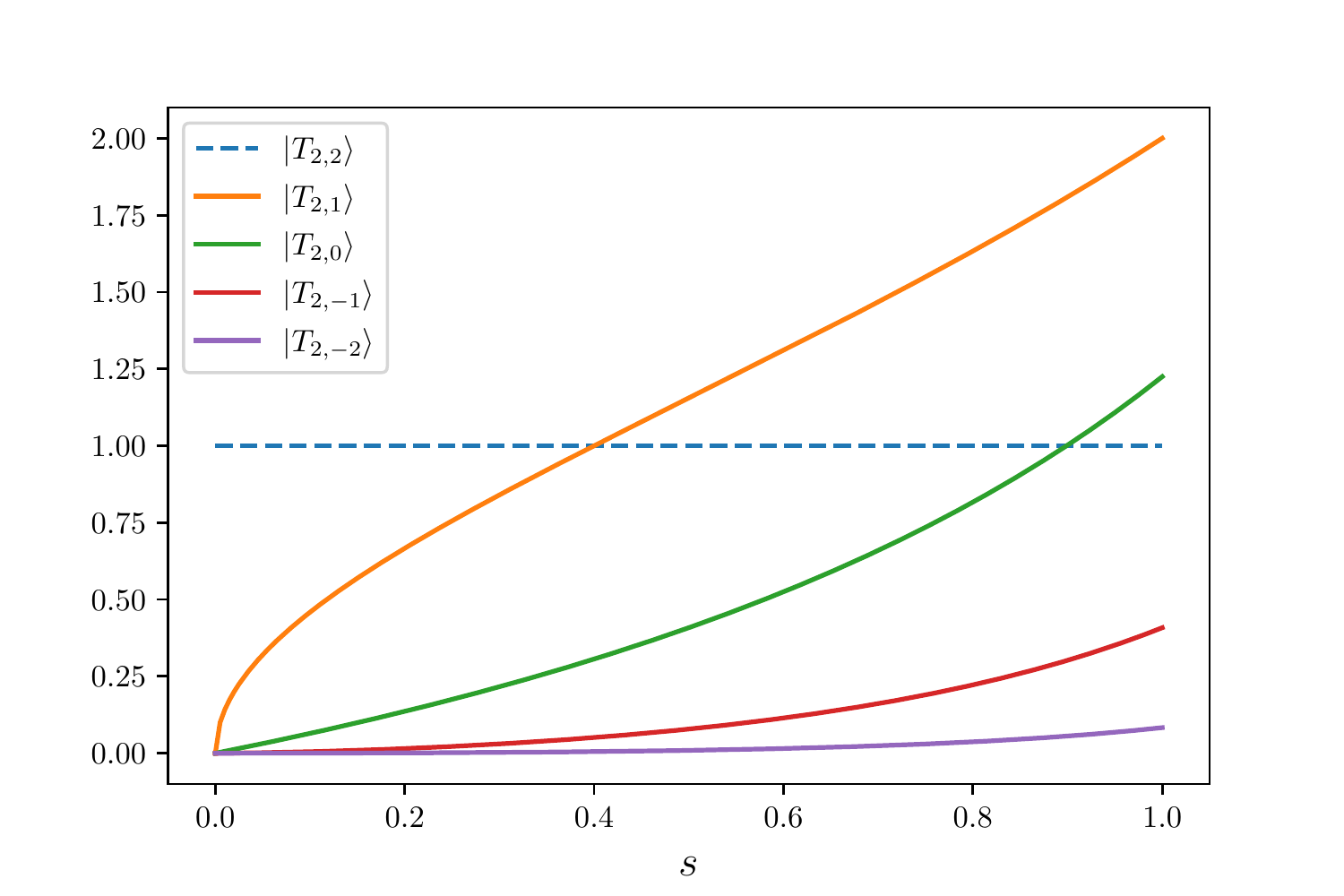}
    \caption{Coefficients of $\ket*{\what{T}_{2,M}} (-2\leq M\leq 2)$ in Eq.~\eqref{eq:|Sn> approximate} as functions of $s$.}
    \label{fig:coeffs}
\end{figure}

\subsection{Estimate of the optimal perturbation strength $\lambda$}
As stated in the main text, we have 
\begin{equation}
    \begin{split}
        (H+\delta H(\lambda))\ket*{S_n}&=P_{\mr{Ryd}}\left(H_{\mr{Z}}+H_{\mr{rem}}(\lambda)\right)\ket*{\wtil{S}_n},\\
        H_{\mr{rem}}(\lambda)&=\sum_{b\in\Lambda_{\mr{B}}}h_{b,b+1}(\lambda),\\
        h_{b,b+1}(\lambda)&=\frac{-1+2\lambda}{\sqrt{6}}\left(\ket*{+,0}+\ket*{0,-}\right)\bra*{T_{2,0}}_{b,b+1}+\frac{\lambda}{\sqrt{2}}\ket*{0,0}\left(\bra*{T_{2,1}}+\bra*{T_{2,-1}}\right)_{b,b+1}.
    \end{split}
\end{equation}
Thus, we find
\begin{equation}
    h_{b,b+1}^\dag(\lambda)h_{b,b+1}(\lambda)=\frac{(-1+2\lambda)^2}{3}\dyad*{T_{2,0}}_{b,b+1}+\frac{\lambda^2}{2}\left(\ket*{T_{2,1}}+\ket*{T_{2,-1}}\right)\left(\bra*{T_{2,1}}+\bra*{T_{2,-1}}\right)_{b,b+1}.
\end{equation}
As $\ket*{T_{2,0}}=\sqrt{3/8}\ket*{\what{T}_{2,2}}-1/2\ket*{\what{T}_{2,0}}+\sqrt{3/8}\ket*{\what{T}_{2,-2}}$ and $\ket*{T_{2,1}}+\ket*{T_{2,-1}}=\ket*{\what{T}_{2,2}}-\ket*{\what{T}_{2,-2}}$, we find
\begin{equation}
    \begin{split}
        \mr{Tr}_{\{b,b+1\}}\rho_Sh^\dag_{b,b+1}(\lambda)h_{b,b+1}(\lambda)&=\frac{(-1+2\lambda)^2}{3}\left(\frac{3}{8}\expval*{\rho_S}{\what{T}_{2,2}}+\frac{1}{4}\expval*{\rho_S}{\what{T}_{2,0}}+\frac{3}{8}\expval*{\rho_S}{\what{T}_{2,-2}}\right)\\
        &+\frac{\lambda^2}{2}\left(\expval*{\rho_S}{\what{T}_{2,2}}+\expval*{\rho_S}{\what{T}_{2,-2}}\right)\\
        &\cong\frac{1}{N}\sum_{s=1/N}^1\left[\frac{(-1+2\lambda)^2}{8Z(s)}\left(1+\frac{s^2}{(2-s)^2}+\frac{s^4}{144(2-s)^4}\right)+\frac{\lambda^2}{2Z(s)}\left(1+\frac{s^4}{144(2-s)^4}\right)\right]\\
        &\cong\int_0^1ds\left[\frac{(-1+2\lambda)^2}{8Z(s)}\left(1+\frac{s^2}{(2-s)^2}+\frac{s^4}{144(2-s)^4}\right)+\frac{\lambda^2}{2Z(s)}\left(1+\frac{s^4}{144(2-s)^4}\right)\right],
    \end{split}
\end{equation}
where $\mr{Tr}_{\{b,b+1\}}$ is the partial trace over the degrees of freedom on $\{b,b+1\}$. This function becomes smallest at $\lambda=\alpha/(2(\alpha+\beta))$, where
\begin{equation}\label{eq:alpha, beta}
    \begin{split}
        \alpha&\coloneqq\int_0^1ds\frac{1}{Z(s)}\left(1+\frac{s^2}{(2-s)^2}+\frac{s^4}{144(2-s)^4}\right)\cong0.5350,\\
        \beta&\coloneqq\int_0^1ds\frac{1}{Z(s)}\left(1+\frac{s^4}{144(2-s)^4}\right)\cong0.4739.
    \end{split}
\end{equation}
Thus, we find the optimal coefficient $\lambda\cong0.2651$. We define integrands in Eq.~\eqref{eq:alpha, beta} as $\alpha(s)$ and $\beta(s)$, i.e., $\alpha=\int_0^1ds\alpha(s), \beta=\int_0^1ds\beta(s)$.